\pdfoutput=1

\documentclass[11pt, reqno,preprint]{article}
\usepackage{jheppub}
\usepackage{epsfig}
\usepackage{amssymb}
\usepackage{amsmath}
\usepackage{mathrsfs}
\usepackage{hyperref}
\usepackage{multirow}

\def\be{\begin{equation}}
\def\ee{\end{equation}}
\def\ba{\begin{eqnarray}}
\def\ea{\end{eqnarray}}
\def\nl{\nonumber\\}

\def\Li{\textrm{Li}}

\def\GG{\mathcal{G}}

\def\zb{{\bar{z}}}
\def\rhob{\bar{\rho}}
\def\hb{{\bar{h}}}

\newcommand{\lsim}{\mathrel{\hbox{\rlap{\lower.55ex \hbox{$\sim$}} \kern-.3em \raise.4ex \hbox{$<$}}}}
\newcommand{\gsim}{\mathrel{\hbox{\rlap{\lower.55ex \hbox{$\sim$}} \kern-.3em \raise.4ex \hbox{$>$}}}}

\def\firstc#1#2{{\tilde{c}(#1,#2)}}
\def\cc#1#2{{c(#1,#2)}}
\def\avg#1{\big\langle #1\big\rangle}

\title{Gravitational S-matrix from CFT dispersion relations}
\author{Luis F. Alday$^{a}$, Simon Caron-Huot$^{b}$}
\affiliation[a]{Mathematical Institute, University of Oxford,
Andrew Wiles Building, Radcliffe Observatory Quarter,
Woodstock Road, Oxford, OX2 6GG, UK }
\affiliation[b]{Department of Physics, McGill University, 3600 rue University, Montreal, QC Canada H3A 2T8}
\emailAdd{alday@maths.ox.ac.uk, schuot@physics.mcgill.ca}
\abstract{
We analyse the double-discontinuities of the four-point correlator of the stress-tensor multiplet in N=4 SYM at large t' Hooft coupling and at order $1/N^4$,
as a way to access one-loop effects in the dual supergravity theory.
From these singularities we extract CFT-data by using two inversion procedures: one based on a recently proposed Froissart-Gribov inversion integral, and the other based on large spin perturbation theory.
Both procedures lead to the same results and are shown to be equivalent more generally. Our computation parallels the standard S-matrix reconstruction via dispersion relations. In a suitable limit, the result of the conformal field theory calculation is compared with the one-loop graviton scattering
amplitude in ten-dimensional IIB supergravity in flat space, finding perfect agreement.
}
\begin{document}

\maketitle

\def\x#1{x_{#1}^2}
\def\xsq#1{x_{#1}^4}
\def\y#1{y_{#1}^2}
\def\a{\alpha}
\def\ab{\overline{\alpha}}
\def\j{J}

\newpage

\section{Introduction}

The gauge-gravity correspondence equates certain gravitational theories to strongly coupled field theories
in a lower dimensional spacetime, enabling a deeper understanding of both.
Strongly coupled field theories are notoriously hard to study directly, but it has been proposed that in the presence of
conformal symmetry they can be quantitatively ``bootstrapped'' by
exploiting symmetry and physical consistency principles \cite{Rattazzi:2008pe}.
This makes it increasingly feasible to look ``under the hood'' of the duality
and, following the program initiated in \cite{Heemskerk:2009pn},
to develop CFT techniques to tackle problems in quantum gravity.

A key realization of the bootstrap program is that the singularities of correlators in a certain limit
determine the large-spin behavior in the expansion around other limits \cite{Fitzpatrick:2012yx,Komargodski:2012ek,Alday:2013cwa}.
Crucially, this works to all orders in an expansion in inverse spin \cite{Alday:2016njk}, see also \cite{Alday:2015ewa,Simmons-Duffin:2016wlq},
thus allowing in principle the full OPE data to be reconstructed just from singular terms.
When applied to large-$N$ theories, it turns out that, with the correct understanding of ``singular'', reviewed below, only \emph{single-trace}
exchanges contribute, making this procedure particularly efficient.
The part of the correlator which is analytic in spin is fully reconstructed \cite{Alday:2016njk,Simmons-Duffin:2016wlq,
Alday:2017gde,Kulaxizi:2017ixa,Li:2017lmh}, while any possible ambiguity is limited to the
lowest few spins ($\ell< 2$) and is otherwise constrained by crossing \cite{Alday:2017gde,Caron-Huot:2017vep}.
The idea of building up all correlators from just the singular part of single-trace exchanges is especially appealing
in theories which have
a low-energy gravity dual, since the spectrum then contains a limited number of single-trace primary operators,
such as the stress tensor dual to the graviton.

The aim of this paper is to carry out this program explicitly in a concrete example of a fully-fledged, consistent CFT. We will show how to explicitly reconstruct, from singularities, all CFT-data in $\mathcal{N}=4$ super Yang-Mills with gauge group $SU(N)$, to order $1/N^4$ at strong `t Hooft coupling.
This corresponds to one-loop in the dual gravitational theory, namely type IIB superstring theory on AdS${}_5\times$S${}_5$.
Having such a specific example at hands makes it possible to ask a variety of sharp questions.
In particular, we will analyze in detail how the CFT correlator encodes the flat space $S$-matrix of 10D supergravity,
at both tree-level and one-loop. This will reveal a direct connection between the method of CFT analyticity and the
traditional method of S-matrix unitarity and dispersion relations.
While we focus our attention on this specific example, we expect the method and philosophy to apply more widely. 

This paper is organized as follows. In section \ref{generalities} we explain the general idea of the method and summarise our results, without entering into technicalities. In section \ref{inversion} we show how to recover the CFT-data from the double-discontinuities of the correlator, using the Froissart-Gribov inversion integral derived in \cite{Caron-Huot:2017vep}.
In section \ref{LSPT} these results are recovered from all order perturbation around large spin, following  \cite{Alday:2016njk}. It is furthermore shown that the two inversion procedures are actually equivalent. In section \ref{flat} we study a particular limit of the CFT correlator which reproduces the S-matrix of the higher-dimensional bulk theory. 
Remarkably, it is found that the limit of the discontinuity of the CFT-correlator reproduces the discontinuity of the bulk theory S-matrix. This allows to draw a perfect parallel between our inversion procedure in CFT and the standard S-matrix reconstruction via dispersion relations.  We also explain in detail the precise ambiguities when following these procedures. With then end up with some conclusions, while some of the technical details are deferred to the appendices. 

\section{Generalities}
\label{generalities}
The stress tensor of ${\cal N}=4$ SYM lies in a short multiplet. The super conformal primary of this multiplet is a scalar operator ${\cal O}_2$, of protected dimension $2$ and which transforms in the $[0,2,0]$ representation of the $SU(4)$ $R-$symmetry group. We will consider the correlator of four identical such operators
\begin{equation}
\langle {\cal O}_2(x_1) {\cal O}_2(x_2){\cal O}_2(x_3){\cal O}_2(x_4) \rangle = \sum_{\cal R} \frac{\GG^{({\cal R})}(u,v)}{x_{12}^4 x_{34}^4}
\end{equation}
where the sum runs over the six representations present in the tensor product $[0,2,0] \times [0,2,0]$ and we have introduced the cross-ratios
\begin{equation}
u=\frac{x_{12}^2x_{34}^2}{x_{13}^2x_{24}^2},~~~v=\frac{x_{14}^2x_{23}^2}{x_{13}^2x_{24}^2}
\end{equation}
The superconformal Ward identities allow to write all functions $\GG^{({\cal R})}(u,v)$ in terms of a single unprotected function $\GG(u,v)$,
equal to $\GG^{(105)}/u^2$, which satisfies the following crossing relation
\begin{equation}
\label{crossing}
v^2 \GG(u,v) - u^2 \GG(v,u) + (u^2-v^2) + \frac{u-v}{c} =0
\end{equation}
with the central charge $c=\frac{N^2-1}{4}$. See \cite{Beem:2016wfs} for the details.%
\footnote{We normalize $\GG$ so that its disconnected part is $1+\frac{1}{v^2}$. Compared with \cite{Beem:2016wfs}, $\GG^{\rm there}=4\GG^{\rm here}$.}
$ \GG(u,v) $ admits the following decomposition
\begin{equation}
\label{Gdec}
\GG(u,v) = \GG^{(short)}(u,v) + {\cal H}(u,v)
\end{equation}
Here $\GG^{(short)}(u,v) $ encodes the contribution from protected intermediate operators, belonging to short multiplets. It does not depend on the coupling constant and has been explicitly computed in \cite{Beem:2016wfs}. ${\cal H}(u,v)$ encodes the dynamical contribution from long multiplets. It admits a decomposition in super-conformal blocks  
\begin{equation}
{\cal H}(u,v) = \sum_{\Delta,\ell} a_{\Delta,\ell} \,g_{\Delta,\ell}(u,v) \label{OPE}
\end{equation}
where the sum runs over super conformal primary operators in long multiplets, with dimension $\Delta$ and spin $\ell$.
Each contribution is weighted by the squared OPE coefficient $a_{\Delta,\ell}$ and the function $g$,
which includes a normalization factor which will be convenient in this work:
\be
 g_{\Delta,\ell}(u,v) =  r_{\frac{\Delta+4+\ell}{2}}r_{\frac{\Delta-\ell+2}{2}} \times u^{-2}\,\tilde{g}_{\Delta+4,\ell}(u,v), \qquad r_h = \frac{\Gamma(h)^2}{\Gamma(2h-1)}\,.
 \label{normalization}
\ee
The standard four-dimensional block $\tilde{g}_{\Delta,\ell}(u,v)$
is in turn expressed most simply in terms of cross ratios $z,\zb$, such that $u=z \zb,v=(1-z)(1-\zb)$:
\be \label{block_4d}
 \tilde{g}_{\Delta,\ell}(z,\zb) =
 \frac{z\zb}{\zb-z} \left[k_{\frac{\Delta-\ell-2}{2}}(z)k_{\frac{\Delta+\ell}{2}}(\zb)-k_{\frac{\Delta+\ell}{2}}(z)k_{\frac{\Delta-\ell-2}{2}}(\zb)\right],
\ee
where, finally, $k_h(z)=z^{h}\,{}_2F_1(h,h,2h,z)$ is a standard hypergeometric function.

Plugging the decomposition (\ref{Gdec}) into (\ref{crossing}) results in a crossing equation for ${\cal H}(u,v)$,
\begin{equation*}
{\cal H}(u,v)+ \GG^{(short)}(u,v) =\frac{u^2}{v^2} {\cal H}(v,u)  +\frac{u^2}{v^2} \GG^{(short)}(v,u)  +  \frac{v^2-u^2}{v^2} + \frac{v-u}{v^2}\,\frac{1}{c}\,.
\end{equation*}
From the explicit result for $\GG^{(short)}(u,v)$ we find the following behaviour for small $v$
\begin{eqnarray}
\label{crossdiv}
{\cal H}(u,v) + \GG^{(short)}(u,v)  &=& \frac{u^2}{v^2} {\cal H}(v,u) + \frac{1}{v^2} +\frac{2 u^2 \log u-3 u^2+4 u-1}{v  (u-1)^3} \,\frac{1}{c}+ \text{regular}
\end{eqnarray}
where the regular terms contain at most a single logarithm as $v\to 0$, in contrast with terms which we will call ``singular'' due to either poles or double logarithms
as $v\to 0$.
So far the discussion has been general. In the present paper we will consider solutions consistent with crossing in a large central charge expansion, in the regime of large t' Hooft coupling $\lambda$:
\begin{equation}
{\cal H}(u,v)= {\cal H}^{(0)}(u,v)+\frac{1}{c} {\cal H}^{(1)}(u,v)+ \frac{1}{c^2} {\cal H}^{(2)}(u,v)+\cdots
\end{equation}
In this regime the intermediate operators contributing to ${\cal H}(u,v)$ are double trace operators with twist four and higher. The poles at $v = 0$ present on the r.h.s. of (\ref{crossdiv}) arise from the protected, single-trace sector.
Following general arguments, we see that these poles are consistent with, and actually require, the existence of double trace operators of twist $\Delta-\ell=4+2n$. As we will see, their precise form at $c=\infty$ suffices to fix the OPE coefficients to
\begin{eqnarray}\label{a0}
\avg{a^{(0)}}_{n,\ell}= 2(\ell+1)(6+\ell+2n)\,.
\end{eqnarray}
We use the bracket to denote the sum over all operators of approximate twist $4+2n$ and spin $\ell$,
emphasizing the fact that in general many nearly-degenerate operators contribute.
As we take into account $1/c$ corrections both the scaling dimensions and OPE coefficients of individual operators acquire corrections
\begin{eqnarray}
\Delta_{n,\ell} &=& 4+2n+\ell+ \frac{1}{c} \gamma^{(1)}_{n,\ell} +  \frac{1}{c^2} \gamma^{(2)}_{n,\ell} + \cdots\\
a_{n,\ell} &=& a^{(0)}_{n,\ell}+ \frac{1}{c}a^{(1)}_{n,\ell} + \frac{1}{c^2}a^{(2)}_{n,\ell} + \cdots
\end{eqnarray}
As we will see in the next two sections $\gamma^{(1)}_{n,\ell} $ and $a^{(1)}_{n,\ell}$ are again fully determined by the singular terms
in (\ref{crossdiv}). We obtain
\be
\frac{\avg{a^{(0)}\gamma^{(1)}}_{n,\ell}}{\avg{a^{(0)}}_{n,\ell}}
= -\frac{\kappa_n}{(1+\ell)(6+\ell+2n)},
\qquad \avg{a^{(1)}}_{n,\ell} = \frac{1}{2} \partial_n\, \avg{a^{(0)}\gamma^{(1)}}_{n,\ell}\,,
\ee
where $\kappa_n=(n+1)(n+2)(n+3)(n+4)$. This coincides with the well known supergravity result. In principle one could also add a solution consistent with crossing with finite support in the spin. As we will show, such solutions can be forbidden using bounds on the Regge limit behavior.

\begin{figure}
 \centering
 \def\svgwidth{0.9\textwidth}
\begingroup%
  \makeatletter%
  \providecommand\color[2][]{%
    \errmessage{(Inkscape) Color is used for the text in Inkscape, but the package 'color.sty' is not loaded}%
    \renewcommand\color[2][]{}%
  }%
  \providecommand\transparent[1]{%
    \errmessage{(Inkscape) Transparency is used (non-zero) for the text in Inkscape, but the package 'transparent.sty' is not loaded}%
    \renewcommand\transparent[1]{}%
  }%
  \providecommand\rotatebox[2]{#2}%
  \ifx\svgwidth\undefined%
    \setlength{\unitlength}{554.54502008bp}%
    \ifx\svgscale\undefined%
      \relax%
    \else%
      \setlength{\unitlength}{\unitlength * \real{\svgscale}}%
    \fi%
  \else%
    \setlength{\unitlength}{\svgwidth}%
  \fi%
  \global\let\svgwidth\undefined%
  \global\let\svgscale\undefined%
  \makeatother%
  \begin{picture}(1,0.24803731)%
    \put(0,0){\includegraphics[width=\unitlength,page=1]{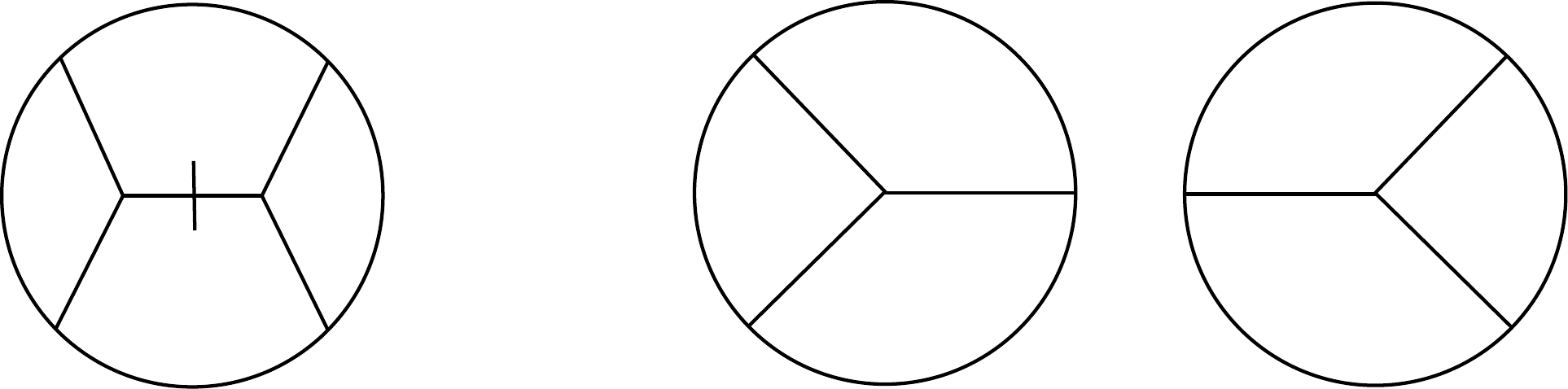}}%
    \put(0.2826104,0.11621766){\color[rgb]{0,0,0}\makebox(0,0)[lb]{\smash{\scalebox{2}{$= \sum$}}}}%
    \put(0.70134795,0.11280219){\color[rgb]{0,0,0}\makebox(0,0)[lb]{\smash{\scalebox{2}{$\times$}}}}%
    \put(0.2518711,0.06908409){\color[rgb]{0,0,0}\makebox(0,0)[lb]{\smash{}}}%
    \put(0.24025845,0.01307023){\color[rgb]{0,0,0}\makebox(0,0)[lb]{\smash{}}}%
    \put(0.3119835,0.04654195){\color[rgb]{0,0,0}\makebox(0,0)[lb]{\smash{{\begin{tabular}{c}\Large single\\\Large traces\end{tabular}}}}}%
  \end{picture}%
\endgroup%

\caption{
Correlators in any large-$N$ theories can be fully reconstructed from singularities (denoted by the cut) that are saturated by single-trace operators.
Theories with a gravity dual correspond to the case where the sum is effectively finite.
}
\label{fig:tree}
\end{figure}
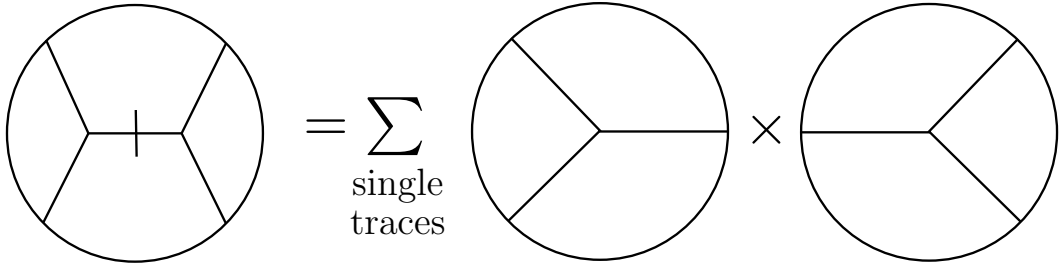

Although in this paper we will only focus on the correlator at hand, in principle the same can be done for more general correlators, of the form $\langle {\cal O}_p{\cal O}_p{\cal O}_q{\cal O}_q\rangle$, which on the gravity side are interpreted as different S${}_5$ spherical harmonics.
In this way one should recover the full supergravity result from the singular contribution of the protected sector.  This is a manifestation of a more general result: correlators in large-$N$ CFTs can be reconstructed from singularities, due to single trace operators, see figure \ref{fig:tree}

\begin{figure}
 \centering
 \def\svgwidth{0.9\textwidth}
\begingroup%
  \makeatletter%
  \providecommand\color[2][]{%
    \errmessage{(Inkscape) Color is used for the text in Inkscape, but the package 'color.sty' is not loaded}%
    \renewcommand\color[2][]{}%
  }%
  \providecommand\transparent[1]{%
    \errmessage{(Inkscape) Transparency is used (non-zero) for the text in Inkscape, but the package 'transparent.sty' is not loaded}%
    \renewcommand\transparent[1]{}%
  }%
  \providecommand\rotatebox[2]{#2}%
  \ifx\svgwidth\undefined%
    \setlength{\unitlength}{516.59864081bp}%
    \ifx\svgscale\undefined%
      \relax%
    \else%
      \setlength{\unitlength}{\unitlength * \real{\svgscale}}%
    \fi%
  \else%
    \setlength{\unitlength}{\svgwidth}%
  \fi%
  \global\let\svgwidth\undefined%
  \global\let\svgscale\undefined%
  \makeatother%
  \begin{picture}(1,0.25876903)%
    \put(0,0){\includegraphics[width=\unitlength,page=1]{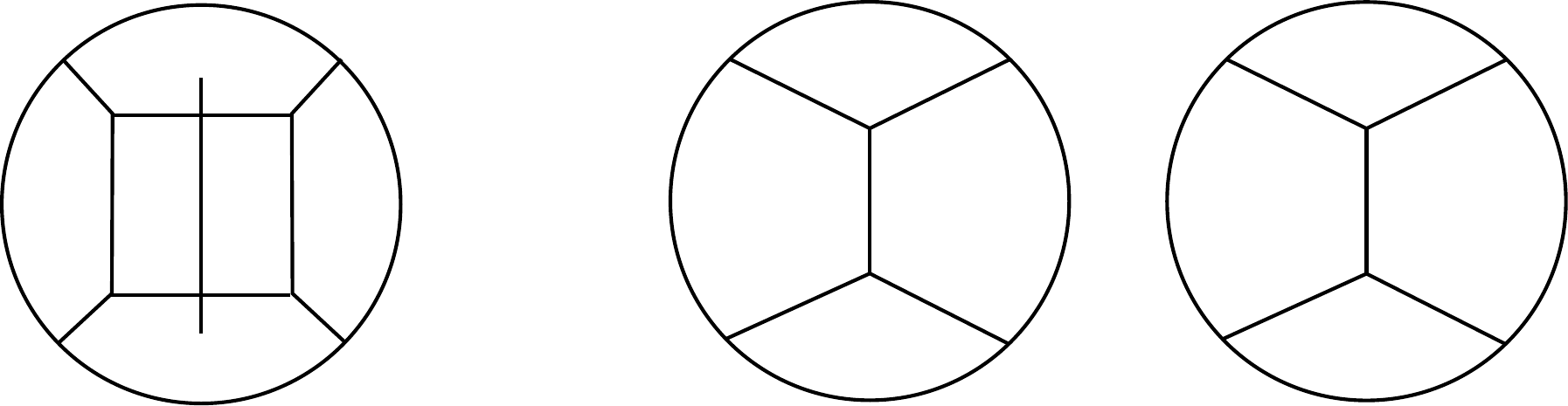}}%
    \put(0.69406592,0.12151705){\color[rgb]{0,0,0}\makebox(0,0)[lb]{\smash{\scalebox{2}{$\times$}}}}%
    \put(0.27879132,0.11481337){\color[rgb]{0,0,0}\makebox(0,0)[lb]{\smash{\scalebox{2}{$= \sum$}}}}%
    \put(0.31032203,0.04001966){\color[rgb]{0,0,0}\makebox(0,0)[lb]{\smash{{\begin{tabular}{c}\Large single\\\Large traces\end{tabular}}}}}%
  \end{picture}%
\endgroup%

\caption{At one-loop order in the $1/N$ expansion, the singularities caused by double-trace exchanges are equal to products
of single-trace tree amplitudes.
}
\label{fig:loop}
\end{figure}
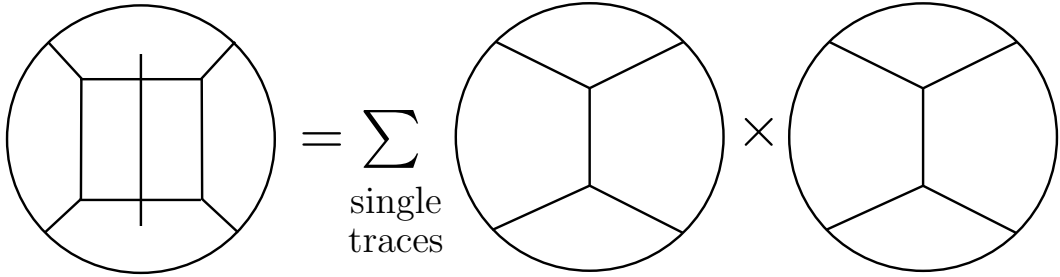

At order $1/c^2$ something interesting happens. On one hand the $1/c$ expansion of the protected contribution stops at order $1/c$. On the other hand, the anomalous dimensions at order $1/c$ of double trace operators in the $t$-channel
produce the following singular term to order $1/c^2$:
\begin{eqnarray} \label{one_loop_log2}
{\cal H}^{(2)}(v,u) = \frac{1}{8} \log^2 v \sum_{n,\ell} \avg{a^{(0)}\big(\gamma^{(1)} \big)^2}_{n,\ell}  \,g_{n,\ell}(v,u) + \text{regular}
\end{eqnarray}
In order to compute these sums one has to solve a mixing problem which can be done from the explicit answers of the general correlators mentioned above. The result is recorded in eq.~(\ref{one-loop gamma squared}) below.
As we show in sections \ref{inversion} and \ref{LSPT} the CFT data at order $1/c^2$ again follows from this singular part. This is shown pictorially in figure \ref{fig:loop}.

According to the AdS/CFT dictionary, this $1/c^2\sim 1/N^4$ CFT correction describes one-loop effects
in the dual gravitational theory.  To demonstrate this more explicitly, in section \ref{flat} we take the limit of large scaling dimension,
where the CFT-data is expected to encode the scattering amplitude of a pair of excitations with large center-of-mass energy in AdS units $\sqrt{s}L=2n$,
through a simple relation \cite{Cornalba:2007zb,Heemskerk:2009pn} (discussed further in section \ref{flat}):
\be
 \lim_{n\to\infty} \frac{\avg{a e^{-i\pi\gamma}}_{n,\ell}}{\avg{a^{(0)}}_{n,\ell}} = b_\ell(s)\,.
\ee
Here $b_\ell(s)$ are the angular momentum partial waves of the flat-space superstring S-matrix,
and $n$ should be large but not too large so that we are still in the regime controlled by supergravity.
We will show that the two sides of this equation agree precisely, revealing how
the operator mixing just mentioned successfully accounts for the ten-dimensional nature of the AdS${}_5\times$S${}_5$ geometry.

\section{CFT data from the Froissart-Gribov inversion integral}
\label{inversion}

Recently, an integral formula has been derived which reconstructs the OPE data of any CFT
from the double-discontinuity of correlators \cite{Caron-Huot:2017vep}.
For identical external operators in four dimensions, this inversion integral was written in that paper
as
\be
 \firstc{\ell}{\Delta} = \frac{1+(-1)^\ell}{4}  \,\tilde{\kappa}(\tfrac{\Delta+\ell}{2})
 \int_0^1 \frac{dz}{z^2}\,\frac{d\zb}{\zb^2}
 \left(\frac{\zb-z}{z\zb}\right)^2 \tilde{g}_{\ell+3,\Delta-3}(z,\zb)\,{\rm dDisc}\,[\GG(z,\zb)],
\ee
with $\tilde{\kappa}(h)=\frac{\Gamma(h)^4}{2\pi^2\Gamma(2h-1)\Gamma(2h)}$,
and where we notice that the block has spin and dimension interchanged compared to the one which enters the OPE above.
The formula is analytic in spin except for the $(-1)^\ell$ prefactor, which we'll now set to 1
since all the exchanged operators have even spin.

The double-discontinuity is defined as the expectation value of a squared commutator in real Minkowski spacetime.
Alternatively, it can be computed as the difference between the Euclidean correlator
and its two possible analytic continuations around $\zb=1$:
\be
{\rm dDisc}\,[\GG(z,\zb)] \equiv \GG(z,\zb) -\tfrac12 \GG^\circlearrowleft(z,\zb)-\tfrac12 \GG^\circlearrowright(z,\zb). \label{ddisc}
\ee
For the one-loop correlator in supergravity, the double-discontinuity is thus simply the coefficient of $\log^2v$, times $4\pi^2$.
At tree-level, the coefficient of $\log^2v$ vanishes, and the double-discontinuity comes
exclusively from the polar terms as $\zb\to 1$, caused by protected single-trace operators.
Thus the double-discontinuity picks up precisely those terms dubbed ``singular'' in the preceding section.

The function $\firstc{\ell}{\Delta}$ encodes the $s$-channel OPE data through its poles \cite{Costa:2012cb,Caron-Huot:2017vep}:
if $\Delta_k$ is the dimension of the exchanged operator, then 
\be
\firstc{\ell}{\Delta}\to \frac{f^2_{\ell,\Delta_k}}{\Delta_k-\Delta}. \label{residue}
\ee
where $f^2_{\ell,\Delta_k} \sim a_{\Delta_k,\ell}$. It will be convenient to switch variables to $h=\frac{\Delta-\ell+2}{2}$, $\hb=\frac{\Delta+\ell+4}{2}$,
in terms of which the inversion integral nicely factorizes.  Using the explicit form of the blocks,
including the normalization in (\ref{normalization}) and the corresponding shift of $\Delta$ by 4, and symmetry in $(z,\zb)$, the
function which extracts the coefficients $a_{\Delta,\ell}$ is given by the integral:
\be
 \cc{h}{\hb} = \int_0^1 \frac{dz}{z^2} \frac{k_{1-h}(z)}{r_{h}}\,
 \int_0^1 \frac{d\zb}{\zb^2}  \frac{r_\hb}{\hb-\frac12}k_{\hb}(\zb) \frac{{\rm dDisc}\,[z\zb(\zb-z)\GG(z,\zb)]}{4\pi^2}\,.
\label{fg}
\ee
This will provide the starting point for the applications in this section.
Since according to eq.~(\ref{residue}) a given double trace operator produces a pole at $h=3+n+\gamma/2$ and $\hb=4+n+\ell+\gamma/2$,
upon summing over nearly-degenerate operators one finds, close to integer $h$:
\be
 \cc{h}{h+\ell+1} = \left\langle \frac{a}{2\big(3+n+\frac12\gamma-h\,\big)}\right\rangle_{n,\ell}. \label{residues}
\ee
Again the bracket stands for the sum over all (superconformal primary) operators with spin $\ell$ and approximate twist $4+2n$
(not necessarily restricted to double trace operators).
Expanding in the anomalous dimension $\gamma\sim 1/c$, one sees that
a single pole around integer $h$ encodes the OPE coefficient, a double pole encodes the averaged anomalous dimension,
a triple pole encodes anomalous dimensions squared, and so on.

In practice, the inversion integral (\ref{fg}) works as follows: the OPE data is encoded in poles with respect to $h$,
which come from the $z\to 0$ limit of integration. Thus different powers of $z$ in that limit yield different twists.
The $\zb$ integral, on the other hand, is dual to $\hb$ and provides the spin dependence for each twist.

\subsection{Computation of integrals: tree-level supergravity}

To illustrate this formula in practice, let us immediately apply it to
the singular terms explicited in eq.~(\ref{crossdiv}), and obtain the tree-level supergravity data.
When converted to $z,\zb$ variables, these singular terms read, for disconnected and connected ($1/c$) contributions:
\begin{subequations}
\label{dDisc01}
\ba
\label{dDisc0}
 {\rm dDisc}\,\left[z\zb(\zb-z)\GG^{(0)}(z,\zb)\right]&=&
 \frac{z}{1-z}{\rm dDisc}\,\left[\frac{\zb^2}{(1-\zb)^2}\right]-\frac{z^2}{(1-z)^2}{\rm dDisc}\,\left[\frac{\zb}{1-\zb}\right],\\
 {\rm dDisc}\,\left[z\zb(\zb-z)\GG^{(1)}(z,\zb)\right]&=& \left(\frac{z}{1-z}-\frac{2z^2}{(1-z)^2}-\frac{2z^3\log z}{(1-z)^3}\right){\rm dDisc}\,\left[\frac{\zb}{1-\zb}\right].
\label{dDisc1}
\ea
\end{subequations}
We have retained all terms with a pole at $\zb\to 1$, but dropped everything else.

It is convenient to isolate powers of $\frac{\zb}{1-\zb}$ because they turn out to integrate to a simple
analytic formula for generic exponent $p$, as recorded in eq.~(4.7) of \cite{Caron-Huot:2017vep}.
This can be derived by starting from a standard integral representation of the hypergeometric function:
\be\begin{aligned}
\int_0^1 \frac{d\zb}{\zb^2} \frac{r_\hb}{\hb-\frac12} k_\hb(\zb)\,\frac{1}{4\pi^2}\,{\rm dDisc}\left(\frac{1-\zb}{\zb}\right)^{p}
&=
\frac{\sin(\pi p)^2}{\pi^2} \int_0^1 \frac{d\zb}{\zb} \frac{dv}{1-\zb v}\left( \frac{\zb v (1-v)}{1-\zb v}\right)^{h-1}
\\ &=\frac{1}{\Gamma(-p)^2} \frac{\Gamma(\hb-p-1)}{\Gamma(\hb+p+1)}\,,
\label{Iint}
\end{aligned}\ee
where the second integral was computed by means of a change of variable $\zb \mapsto t=\frac{\zb(1-v)}{1-\zb v}$, which nicely decouples $t$ and $v$.
A multiplicative factor $2\sin(\pi p)^2$ on the first line arose from the double-discontinuity of $(1-\zb)^p$.
Note that for negative integer $p$, this factor produces a double zero which is however canceled by a singularity of the integral, so the terms
in (\ref{dDisc01}) do \emph{not} integrate to zero.
That the formula (\ref{Iint}) can be analytically continued and trusted for negative integer $p$ is clear from the derivation of the
Froissart-Gribov formula in \cite{Caron-Huot:2017vep}, which starts from a nonsingular contour integral.

The $z$ integral can be done using the same change of variable and we find, again for generic $k$:
\ba
 \int_0^1 \frac{dz}{z^2}\frac{k_{1-h}(z)}{r_{h}}\,\left(\frac{z}{1-z}\right)^k
 &=& \frac{\pi\sin(\pi (h+k))\tan(\pi h)}{2\sin(\pi(k-h))\sin^2(\pi k)} \frac{1}{\Gamma(k)^2} \frac{\Gamma(h+k-1)}{\Gamma(h+1-k)}
 \nl
 &\simeq& \pi\cot(\pi(k-h))\times \frac{1}{\Gamma(k)^2}\frac{\Gamma(h+k-1)}{\Gamma(h+1-k)} \label{z_int}
\ea
where on the second line we have dropped terms with no poles near positive integer $h$ (for positive $k$),
since these will not affect the OPE data extracted from the residues.
Physically, the exponent $k$ is (one plus half) the twist of the exchanged operator and in practice is always positive.
The formula produces the poles at $h=k,k+1\ldots$ that one might have expected.%
\footnote{
The integral also gives spurious poles at half-integer $h$ which originate from $k_{1-h}$.
As noted below eq.~(3.9) of \cite{Caron-Huot:2017vep}, these should be canceled by adding a reflected block,
which in our normalizations amounts to symmetrizing in $h\to 1-h$. The OPE data near integer $h$ is unaffected.}

Given the double-discontinuity (\ref{dDisc0}), the inversion integral (\ref{fg}) factorizes and can be done
using the preceding formulas. This gives directly the disconnected OPE data:
\be
 c^{(0)}(h,\hb)= \pi\cot(-\pi h) \big(\hb(\hb-1)-h(h-1)\big). \label{c0_fg}
\ee
At the zeroth order we can neglect the anomalous dimension in eq.~(\ref{residues}) which thus gives
\ba
 \avg{a^{(0)}}_{n,\ell}&=& -2{\rm Res}_{h=n+3} \,c^{(0)}(h,h+\ell+1)
 \nl &=& 2(\ell+1)(6+\ell+2n)
\ea
exactly as quoted in eq.~(\ref{a0}).

For the connected tree (second line of (\ref{dDisc1})) the integral is very similar but there is an extra log.
We can simplify our life somewhat by writing it using Casimir operators:
\be
 \left(\frac{z}{1-z}-\frac{2z^2}{(1-z)^2}-\frac{2z^3\log z}{(1-z)^3}\right) = -\frac12 D(D-2) \frac{z\log z}{1-z}\,,
\qquad D = z^2\partial_z(1-z)\partial_z\,. \label{Casimir}
\ee
The Casimir operators can be integrated by parts and simply give a multiplicative factor equal to
their eigenvalue on the blocks, namely: $(h-2)(h-1)h(h+1)$.
A similar trick will greatly simplify things at one-loop, as shown below.
We do not need to worry about boundary terms in $z$ since poles originate only from $z\to 0$.
To perform the integral over $\frac{z\log z}{1-z}$ we then simply expand it in powers of $z/(1-z)$ and
apply the formula (\ref{z_int}) termwise.  Dropping terms with no poles we obtain a very simple result:
\be
 c^{(1)}(h,\hb) = \frac{\pi^2}{\sin(\pi h)^2} \frac{(h-2)(h-1)h(h+1)}{2}. \label{c1_fg}
\ee
Comparing with (\ref{residues}) then give the (summed) anomalous dimension and OPE coefficient:
\be
 \avg{a^{(0)}\gamma^{(1)}}_{n,\ell} = -2\kappa_n,\qquad \avg{a^{(1)}}_{n,\ell}=-\partial_n \kappa_n
\ee
which again are in precise agreement with the results quoted in the preceding section.

\begin{figure}
 \centering
 \def\svgwidth{0.9\textwidth}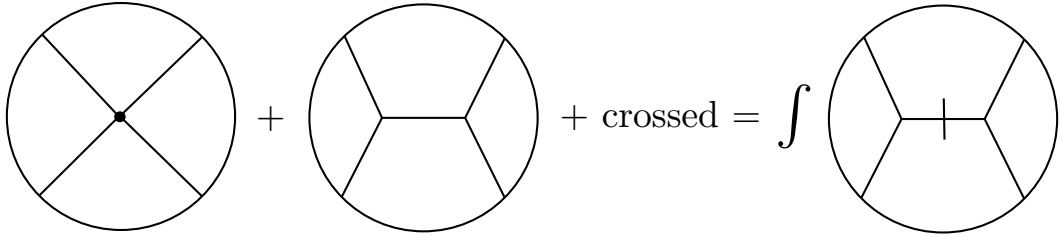
\caption{
The inversion integral produces the full correlator, given on the left as a sum over Witten diagrams,
from the double-discontinuity in a single channel (and more generally, the $t$ and $u$ channels).
}
\label{fig:tree1}
\end{figure}

Let us now interpret the results (\ref{c0_fg}), (\ref{c1_fg}).
They express the result of the inversion integral (\ref{fg}) applied to strongly coupled super Yang-Mills theory
(where one has only neglected terms with no poles at positive $h$).
We stress that this data determines the \emph{full} tree-level supergravity correlator: plugging the resulting anomalous dimensions
and OPE coefficients into eq.~(\ref{OPE})
we checked that it reproduces precisely the OPE expansion of the known result \cite{Arutyunov:2000py}:
\be
 \mathcal{G} = 1 + \frac{1}{v^2} + \frac{1}{c} \left(\frac{1}{v} - u^2 \bar{D}_{2,4,2,2}(z,\zb)\right) +O(1/c^2). \label{Gtree}
\ee
It is remarkable that the present computation did not use any input from supergravity: the only assumption was the sparseness of the single-trace spectrum.
Specifically, we included in the $t$-channel only the protected half-BPS operators
(the stress tensor multiplet and its second Kaluza-Klein excitation), which are responsible for the singular part of $G^{\rm short}(v,u)$ recorded
in eq.~(\ref{crossdiv}).

Physically, this can be viewed as a Kramers-Kronig-type dispersion relation in the bulk of AdS, as represented pictorially in fig.~\ref{fig:tree1}.
The discontinuity produced by single-trace operators is drawn as a cut diagram.
The inversion integral reconstructs the full OPE data from this absorptive contribution.

As a technical comment, we note that the precise form of $\mathcal{G}^{\rm short}$ was never needed on the left-hand-side of eq.~(\ref{crossdiv}),
only its singular part on the right-hand-side. This is because $\GG^{\rm short}(u,v)$ contains only twists below the double-trace threshold,
and in the inversion formula these only lead to poles at $h\leq 2$ which do not contaminate the unprotected data at $h\geq 3$.

Reconstructions via dispersion relations typically suffer from polynomial ambiguities.
This is reflected in the well-known fact that the Froissart-Gribov integral may fail for a finite number
of low spins, equal to the order of the ambiguity.
In a full, unitary CFT, this can only affect $\ell=0,1$, as shown in \cite{Caron-Huot:2017vep}.
Order-by-order in a perturbative $1/c$ expansion, however, the situation can be worse, and a larger (but still finite)
number of low spins can be affected.
Such ambiguities are still constrained by crossing and thus correspond one-to-one to higher-dimensional operators in the bulk effective Lagrangian \cite{Heemskerk:2009pn}.

The fact that the tree-level supergravity result is determined, up to these ambiguities, by the singularities caused by half-BPS single-trace operators
has been understood for some time, see \cite{Dolan:2006ec,Penedones:2010ue} and most recently \cite{Rastelli:2016nze,Rastelli:2017udc,Aprile:2017bgs}.
A chief advantage of the present methods are that it is not necessary to make any a-priori assumption about the space of functions.
In comparison with the Mellin space approach, where unitarity is naturally formulated in terms of very similar pictures (see for example \cite{Fitzpatrick:2012cg}),
the method allows us to extract directly the CFT-data for all $n$, which technically would seem to make the one-loop mixing problem much easier to study.
Furthermore, it is actually possible to bound, using CFT input only, the size of the ambiguities
and show that the solution given above is the \emph{correct} one in the limit of large gap, as
will be discussed in detail below in section \ref{uniqueness}.

\begin{figure}
 \centering
 \def\svgwidth{0.6\textwidth}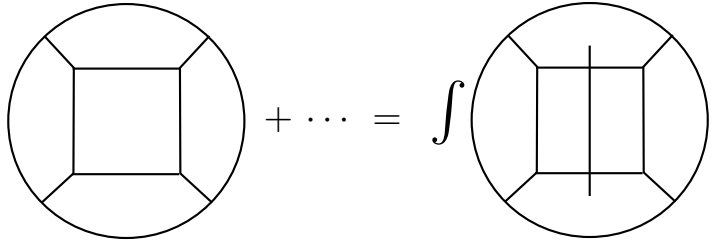
\caption{
The full one-loop correlator, equal to a sum over many Witten diagrams,
is reconstructed from the double-discontinuity in a single channel.
}
\label{fig:loop1}
\end{figure}

\subsection{One loop}

The calculation of the one-loop supergravity correlator is made possible, from the CFT perspective, by the fact
that the double-discontinuity arises from $\log^2$ terms, and therefore solely from the square of tree-level anomalous dimensions.
One only has to ``square'' tree-level data.
As depicted in fig.~\ref{fig:loop1}, this is in perfect match with
the unitarity method in $S$-matrix theory, which bootstraps one-loop amplitudes from products of two trees. For a theory of scalar fields without degeneracy this was used in \cite{Aharony:2016dwx} to compute loops in AdS. In the present case one has to sum over degenerate intermediate states, which here can be pairs of half-BPS operators of arbitrary $R$-charge $p$.
This requires all tree amplitudes of the form $\langle {\cal O}_2{\cal O}_2{\cal O}_p{\cal O}_p\rangle$, not only the $p=2$ case discussed above.
In CFT language, this accounts for operator mixing between different double traces.  This will be studied further with the present methods in \cite{Anh Khoi to appear}.

Our starting point here will be the following compact formula for the sum in \cite{Alday:2017xua}
(the mixing problem in the singlet sector has also been solved in \cite{Aprile:2017bgs} and more generally in \cite{Aprile:2017xsp}):
\be
\frac{\avg{a^{(0)}\big(\gamma^{(1)} \big)^2}_{n,\ell}}{\avg{a^{(0)}}_{n,\ell}} =
\sum_{p=2}^{n+2} \frac{\alpha_p \kappa_n^2}{(J^2-(n+2)(n+3))^2}\prod_{k=2}^{p-1} \frac{(n+2)(n+3)-k(k+1)}{J^2-k(k+1)}
\label{one-loop gamma squared}
\ee
with $\kappa_n=(n+1)(n+2)(n+3)(n+4)$, $J^2=\hb(\hb-1)$ and $\alpha_p=p^2(p^2-1)/12$.

According to the OPE, this determines the coefficient of $\tfrac1{8c^2}\log^2u$, and therefore, using crossing to interchange
the $u$ and $v$ channels, we obtain the double-discontinuity mentioned in eq.~(\ref{one_loop_log2}).
This sum can be performed analytically by making an ansatz in terms of transcendental weight 2 functions of $z$ and $\zb$, multiplied by rational functions
with powers of $(z-\zb)$ in the denominator, and checking that the ansatz matches the expansion to very high order.
The resulting double-discontinuity turns out to be expressible much more simply as the derivative of a ``primitive'':
\be
 \frac{1}{4\pi^2} 
  {\rm dDisc}\,\left[z\zb(\zb-z)\GG^{(2)}(z,\zb)\right] = D(D-2)\bar D(\bar D-2) G^{(2)'}(z,\zb), \label{dDisc2}
\ee
where $G^{(2)'}(z,\zb)$ is recorded in appendix \ref{app:ddisc} and the Casimir operators are given in eq.~(\ref{Casimir}). 
Since the Casimir operators can be integrated by parts (and the primitive vanishes sufficiently fast at $\zb\to 1$ to preclude any boundary term),
the one-loop OPE data is thus written as
\be
 c^{(2)}(h,\hb) = (h-2)(h-1)h(h+1)J^2(J^2-2) \times c^{(2)'}(h,\hb),
\ee
where
\be
c^{(2)'}(h,\hb)\equiv \int_0^1 \frac{dz}{z^2} \frac{k_{1-h}(z)}{r_{h}}\,
 \int_0^1 \frac{d\zb}{\zb^2}  \frac{r_\hb}{\hb-\frac12}k_{\hb}(\zb)\,G^{(2)'}(z,\zb)\,.
\ee
As a simple illustration, let us extract the average of $(\gamma^{(1)})^2$ from the formula.
It comes from the coefficient of $\log^2z$, which is rather compact:
\be
 \label{G2ty}
 G^{(2)'}(z,\zb) = \log^2(t) \frac{t^3y^2(1+ty)}{16(1-t y)^6} =
 \frac{\log^2(t)}{16\times 120}\sum_{q=2}^\infty q(q^2-1)(q+2)(2q+1) t^{q+1}y^{q}
\ee
where $t=\frac{z}{1-z}$ and $y=\frac{1-\zb}{\zb}$. 
The $z$ integral can be performed using (a second derivative of) eq.~(\ref{z_int}) while the $\zb$ integral can be performed using the following
variant of (\ref{Iint}):
\be
 \int_0^1 \frac{d\zb}{\zb^2} \frac{r_\hb}{\hb-\frac12} k_\hb(\zb)\left(\frac{1-\zb}{\zb}\right)^{p}
=\frac{\Gamma(p+1)^2}{2} \frac{\Gamma(\hb-p-1)}{\Gamma(\hb+p+1)}\,. \label{Iint1}
\ee
Performing the sum we find directly
\be
 \avg{a^{(0)}\big(\gamma^{(1)}\big)^2}_{n,\ell} =
 \sum_{q=2}^{n+2} \frac{\kappa_n^2 q(q^2-1)(q+2)(2q+1)}{60\big(q(q+1)-J^2\big)}
 \prod_{k=2}^{q-1} \frac{(n+2)(n+3)-k(k+1)}{J^2-k(k+1)}
\ee
Remarkably, using some telescopic identities, this can be shown to be precisely equivalent to (\ref{one-loop gamma squared}), for all $n$!
Physically, this agreement reflects crossing symmetry of the quadruple discontinuity $\GG^{(2)}\big|_{\log^2u\log^2v}$,
which is a non-obvious but true fact about eq.~(\ref{one-loop gamma squared}). As shown in \cite{Alday:2017xua} under some mild assumptions (\ref{one-loop gamma squared}) is the most general structure consistent with this symmetry. Explicit results can then be used to fix $\alpha_p$. 

To record the new information on the one-loop anomalous dimension and OPE coefficients,
it is useful to expand $c$ around its poles, keeping the value of $\hb$ fixed:
\be
 c^{(2)}(h\to 3+n+\delta/2,\hb) = -\frac{S_{n,\hb}^{(2)}}{8\delta^3}
 -\frac{S_{n,\hb}^{(1)}}{4\delta^2} -\frac{S_{n,\hb}^{(0)}}{\delta}\,.
\label{fixed_h_expansion}
\ee
The poles at fixed $\hb$ are advantageous because they automatically respect the \emph{reciprocity} property:
the large-spin expansion is symmetrical in $\hb\to 1-\hb$.
The fixed-$\ell$ CFT-data is then obtained simply by expanding the $\hb$ dependence in eq.~(\ref{residues}); for example
\ba
 \avg{a^{(0)}\big(\gamma^{(1)} \big)^2}_{n,\ell} &=&  S_{n,\hb}^{(2)} \nl
 \avg{a^{(0)}\gamma^{(2)} + a^{(1)}\gamma^{(1)}}_{n,\ell} &=& S_{n,\hb}^{(1)}
+ \frac12\partial_\hb S_{n,\hb}^{(2)}\label{fixed_hb_averages} \\
\avg{a^{(2)}}_{n,\ell} &=& S_{n,\hb}^{(0)} + \frac12\partial_\hb S_{n,\hb}^{(1)}
+ \frac18 \partial_\hb^2 S_{n,\hb}^{(2)} \nonumber
\ea
where one sets $\hb= n+4+\ell$ after evaluating the derivatives.

\subsection{Results for one-loop anomalous dimensions}

For any desired $n$, the one-loop anomalous dimension part can be obtained
similarly by expanding $\GG^{(2)'}$ to the sufficient order in $\zb$; the result is still a polynomial in $1/z$ so the $z$ integral
can be straightforwardly done using eq.~(\ref{Iint1}).  For the first few twists we find
\ba
S_{0,\hb}^{(1)}&=&\frac{96(16J^2-635)}{(J^2-6)(J^2-20)}\\
S_{1,\hb}^{(1)}&=&\frac{480 (-91710 - 2403 J^2 + 56 J^4)}{(-30 + J^2) (-12 + J^2) (-6 + J^2)}\\
S_{2,\hb}^{(1)}&=&\frac{1440 (-12910968 - 597906 J^2 - 6073 J^4 + 134 J^6)}{(-42 + J^2) (-20 + J^2) (-12 + J^2) (-6 + J^2)}\\
S_{3,\hb}^{(1)}&=&\frac{3360 (-1921913280 - 107519496 J^2 - 2335742 J^4 - 12349 J^6 + 262 J^8)}{(-56 + J^2) (-30 + J^2) (-20 + J^2) (-12 + J^2) (-6 + J^2)}.
\ea
In fact, by series expanding in $t$ and $y$ as in eq.~(\ref{G2ty}) above and keeping also the terms with a single
power of $\log(t)$, we obtained a closed formula valid for all $n$:
\be\begin{aligned}
S_{n,\hb}^{(1)}&= -\frac12\frac{\partial_n \kappa_n}{\kappa_n} S_{n,\hb}^{(2)}
 + \sum_{q=2}^{n+2} \frac{\kappa_n^2 q(q+1)(q+2)}{120(J^2-q(q+1))}
 \left(\prod_{k=2}^{q-1} \frac{(n+2)(n+3)-k(k+1)}{J^2-k(k+1)}\right)\times
\\
& \hspace{-4mm}
 \left[ D_q+ \sum_{r=q+1}^{n+2} \left( \frac{120(q+r+1)(r-1)}{\prod_{m=0}^5(r+m-q)} - \frac{q(q-1)(2q+1)}{r(r-q)}\right)
 \prod_{k'=q}^{r-1} \frac{(n+2)(n+3)-k'(k'+1)}{-k'(k'+1)}\right] \label{S1_alln}
\end{aligned}\ee
where
\ba
D_q&=& \frac{5(q^2-1)(q+3)}{12} \left( \frac{(2q+1)(q-2)}{J^2-(q+1)(q+2)}-\frac{(2q+3)(q+4)}{J^2-(q+2)(q+3)}\right)
\nl && - \frac{60-77q-437q^2+274q^3}{60q} + 2(q-1)(2q+1)\sum_{k'=0}^{q-1}\frac{1}{n+3+k'}\,.
\ea
Possibly this could be further simplified using
telescopic identities as mentioned above, but we have not attempted this.

Let us further discuss the lowest twist $n=0$, where there is no operator mixing.
The above data is then related to the observable anomalous dimension, as follows.
First, we convert the fixed-$\hb$ averages to fixed-$\ell$ averages using eq.~(\ref{fixed_hb_averages}):
\ba
a^{(0)}_{0,\ell} \gamma_{0,\ell}^{(2)} + a^{(1)}_{0,\ell}\gamma^{(1)}_{0,\ell} &=& 
 S_{0,\hb}^{(1)} + \frac12\partial_\hb S_{0,\hb}^{(2)}
\nl &=& \frac{96 (16\ell^4+212\ell^3+311\ell^2-2627\ell-2322)}{(\ell-1)(\ell+1)^2 (\ell+6)^2 (\ell+8)}
\ea
where we have dropped the averaging symbols since there is no mixing.
We then isolate the anomalous dimension by subtracting the product $a_{0,\ell}^{(1)}\gamma_{0,\ell}^{(1)}$,
known from the tree-level computation, giving:
\be
 \gamma^{(2)}_{0,\ell} = \frac{24 (7 \ell^4+ 74 \ell^3-553 \ell^2- 4904 \ell  -3444)}{(\ell-1)(\ell+1)^3(\ell+6)^3(\ell+8)}. \label{gamma2}
\ee
This is the quantity which should be compared with the discrete spectrum (and agrees with eq.~(7.86) of \cite{Aprile:2017bgs}).
This gives, for example, $\gamma_{0,2}^{(2)}=-41/16$ and $\gamma_{0,4}^{(2)}=-423/3125$.
Re-expanding the anomalous dimension $\gamma_{0,\ell}^{(2)}$ in $N$ instead of $c=\frac{N^2-1}{4}$, this agrees
precisely with the $\ell=2,4$ values recorded in \cite{Alday:2017xua}
(namely, $\Delta_{0,2}=6  - 4/N^2- 45/N^4$ and $\Delta_{0,4}=8 - 48/(25 N^2)- 12768/(3125 N^4)$).

Let us briefly comment on the $\ell=1$ pole visible in the preceding formulas. In (\ref{one-loop gamma squared}) each term labelled by $p$ can be interpreted as the contribution from a specific KK-mode to the whole discontinuity and in particular to the CFT-data at order $1/c^2$. The pole at  $\ell=1$ reflects the fact that the $p$-sum diverges for $\ell\leq 1$,
implying that the above formulas are only valid for $\ell\geq 2$.
In principle one could still analytically continue the result (\ref{gamma2}) to $\ell=0$, a procedure which is closely related
(up to a finite counter-term) to throwing away quadratic divergences in the supergravity computation,
as would be automatic using e.g.  dimensional regularization.
However as discussed in section \ref{uniqueness}, the result of such a prescription would violate unitarity and yield
a theory in the ``swampland''.
In reality the true value of $\gamma_{0,0}$ can only be determined nonperturbatively,
and after that, $\avg{a^{(0)}\gamma^{(2)} + a^{(1)}\gamma^{(1)}}_{n,\ell} $ is completely fixed.

One can apply the same method to OPE coefficients $S^{(0)}$.  Explicit results are recorded in appendix \ref{app:ope}.

\section{CFT data from large spin perturbation theory}
\label{LSPT}
In this section we will discuss how to compute the CFT-data from the point of view of large spin pertubation theory, developed in \cite{Alday:2016njk}. After recovering the results of the previous section, we will actually show that the two methods are equivalent. Our starting point is the crossing relation (\ref{crossdiv}). In the regime we are considering the non-protected operators contributing to ${\cal H}(u,v)$ have approximate twist four and higher. As a result ${\cal H}(v,u) \sim v^{2}$ for small $v$ and the divergences explicitly shown on the r.h.s. of (\ref{crossdiv}) have to be reproduced by ${\cal H}(u,v)$. At zeroth order in $1/c$ the arguments of \cite{Fitzpatrick:2012yx,Komargodski:2012ek} imply the existence of towers of large spin operators, of approximate twist $\tau=4+2n$ and OPE coefficients which for large spin behave as
\begin{equation}
\label{zeroOPE}
\avg{a^{(0)}}_{n,\ell}= 2(\ell+1)(6+\ell+2n)
\end{equation}
in the normalization of (\ref{normalization}).
The arguments of \cite{Alday:2016njk} allow us to make a much stronger claim. The behaviour (\ref{zeroOPE}) should be valid to all orders in $1/\ell$. A correction to this behaviour would produce a divergence proportional to $\log^2v$, not present at this order. 

\subsection{Supergravity CFT data}
Next, let us compute the CFT data to order $1/c$. Let us first focus in the anomalous dimension. This should be such that the following divergence is reproduced, see (\ref{crossdiv})
\begin{equation}
\left. {\cal H}^{(1)}(u,v) \right|_{div} = \frac{1}{v} \frac{2 u^2}{(u-1)^3} \log u + \cdots
\end{equation}
From the conformal partial wave decomposition we find
\begin{equation}
{\cal H}^{(1)}(u,v) = \frac{1}{2} \sum_{n,\ell} \avg{a^{(0)}\gamma^{(1)}}_{n,\ell}\, g_{n,\ell}(u,v) \log u + \cdots
\end{equation}
where we have introduced a short-cut notation for the (super)-conformal block with dimension $\Delta=4+2n+\ell$. Following \cite{Alday:2016njk} we will exploit the fact that the anomalous dimension admits an expansion in inverse powers of the conformal spin 
\begin{equation}
\avg{a^{(0)}\gamma^{(1)}}_{n,\ell}= \sum_m \frac{c_{m,n}^{(1)}}{J^{2m}} 
\end{equation}
where $J^2=(n+\ell+3)(n+\ell+4)$, the conformal spin at zeroth order in $1/c$. Hence 
\begin{equation}
\label{TCBdec}
\left. \frac{1}{2} \sum_{n,m} c_{m,n}^{(1)} H_n^{(m)}(u,v)\right|_{div} =  \frac{1}{v} \frac{2 u^2}{(u-1)^3}
\end{equation}
where we have defined the twist conformal blocks (TCB), given by
\begin{equation}
H_n^{(m)}(z,\zb)= \sum_\ell \frac{g_{n,\ell}(z,\zb)}{J^{2m}}\,.
\end{equation}
We will be interested in their divergent contribution as $\zb \to 1$.
This contribution admits the following factorized form 
\begin{equation}
H_n^{(m)}(z,\zb)= \frac{1}{z\zb (\zb-z)}  r_{h_n}k_{h_n}(z) \bar H_n^{(m)}(\zb) + \text{regular}
\end{equation}
where $h_n=\frac{\tau_n+2}{2}=3+n$. From the results in appendix \ref{apTCB} it follows that the functions $\bar H_n^{(m)}(\zb)$ are actually independent of $n$, so that from now on we will drop that index. We are interested in finding the coefficients $c_{m,n}^{(1)}$ in (\ref{TCBdec}).
First let us single out the divergent contribution arising from a specific twist $\tau_n$. This can be done with the help of the following projectors
\begin{equation}
\frac{1}{2\pi i}\oint \frac{dz}{z^2}\, k_{h_n}(z) k_{1-h_{n'}}(z) = \delta_{n,n'}
\end{equation}
Projecting over the contribution of each twist $\tau_n$ we obtain
\begin{equation}
\left. \frac{1}{2} \sum_{m} c_{m,n}^{(1)} \bar H^{(m)}(\zb)\right|_{div} =
\frac{1}{2\pi i} \oint \frac{dz}{z^2} \frac{k_{1-h_n}(z)}{r_{h_n}} z\zb(\zb-z)\,\frac{1}{v} \frac{2 u^2}{(u-1)^3} =
 -\frac{1}{2}\kappa_n \frac{\zb}{1-\zb} + \text{regular}
\end{equation}
As shown in appendix \ref{apTCB} this precise divergence leads to a very simple result, actually independent of the conformal spin. Hence we find 
\begin{equation}
\avg{a^{(0)}\gamma^{(1)}}_{n,\ell}\,=-2 \kappa_n
\end{equation}
Which agrees with the known result. In order to find the corrections to the OPE coefficients we write the full correlator to this order as follows 
\begin{equation}
\label{H1dec}
{\cal H}^{(1)}(z,\zb) = \frac{1}{2} \sum_{m,n} \partial_n \left( c_{m,n}^{(1)}H_n^{(m)}(z,\zb) \right) + \sum_{m,n} d_{m,n}^{(1)}H_n^{(m)}(z,\zb)
\end{equation}
where $d_{m,n}^{(1)}$ are related to the large $J$ expansions of the OPE coefficients as follows
\begin{equation}
\avg{a^{(1)}}_{n,\ell}  =  \sum \frac{d_{m,n}^{(1)}}{J^{2m}}+ \frac{1}{2} \partial_n \left( a_{n,\ell}^{(0)} \gamma_{n,\ell}^{(1)} \right)
\end{equation}
The coefficients $d_{m,n}^{(1)}$ and the expression for the OPE coefficients can be fixed as before, where now
\begin{equation}
\left. \sum_{m,n} d_{m,n}^{(1)}H_n^{(m)}(z,\zb) \right|_{div} = \frac{2 u^2 \log u-3 u^2+4 u-1}{v  (u-1)^3}- \left. \frac{1}{2} \sum_{m,n} \partial_n \left( c_{m,n}^{(1)}H_n^{(m)}(z,\zb) \right)\right|_{div}
\end{equation}
The second term on the r.h.s. contains two contributions. One proportional to $\log u$, which exactly cancels the corresponding piece in the first term, and an extra contribution which can be explicitly computed, since we have already found the coefficients $c_{m,n}^{(1)}$. Once this is done we can proceed as before and fix $d_{m,n}^{(1)}$. We find $d_{m,n}^{(1)}=0$. This leads to the following result for the OPE coefficients
\begin{equation}
\avg{a^{(1)}}_{n,\ell}  = \frac{1}{2} \partial_n \,\avg{a^{(0)}\gamma^{(1)}}_{n,\ell}
\end{equation}
in full agreement with the known result. 

\subsection{One loop}

At next order the correlator admits the following decomposition

\begin{equation}
\label{H2dec}
{\cal H}^{(2)}(z,\zb) = \frac{1}{8} \sum_{m,n} \partial^2_n \left( c_{m,n}^{(2)}H_n^{(m)}(z,\zb) \right)+ \frac{1}{2} \sum_{m,n} \partial_n \left( d_{m,n}^{(2)}H_n^{(m)}(z,\zb) \right) + \sum_{m,n} e_{m,n}^{(2)}H_n^{(m)}(z,\zb)
\end{equation}
where $c_{m,n}^{(2)},d_{m,n}^{(2)},e_{m,n}^{(2)}$ are the coefficients in the large $J$ expansion of specific combinations of CFT-data
\begin{eqnarray}
\label{CFT_data_from_asymptotics}
\sum_m \frac{c_{m,n}^{(2)}}{J^{2m}} &=&  \avg{a^{(0)}\big(\gamma^{(1)}\big)^2}_{n,\ell},  \\
\sum_m \frac{d_{m,n}^{(2)}}{J^{2m}} &=& \avg{a^{(0)}\gamma^{(2)}+a^{(1)}\gamma^{(1)}}_{n,\ell} - \frac{1}{2} \partial_n \,\avg{a^{(0)}\big(\gamma^{(1)}\big)^2}_{n,\ell}, \\
\sum_m \frac{e_{m,n}^{(2)}}{J^{2m}}  &=& \avg{a^{(2)}}_{n,\ell}  - \frac{1}{2}\partial_n \left(\avg{a^{(0)}\gamma^{(2)}+a^{(1)}\gamma^{(1)}}_{n,\ell}\right)
+ \frac{1}{8} \partial^2_n \,\avg{a^{(0)}\big(\gamma^{(1)}\big)^2}_{n,\ell}.
\end{eqnarray}
As before, the angle brackets represent sums over all nearly-degenerate operators. The reciprocity principle implies that these are the combinations that admit an expansion in integer powers of $J^2$, see \cite{Alday:2015eya}. In the relations above weighted averages are shown explicitly. The unknown expansions should be such that the precise divergences are reproduced. As already mentioned in section \ref{generalities} the divergences at this order are proportional to $\log^2 v$.  Let us show in detail how to fix $\langle a_{n,\ell}^{(0)}\left(\gamma_{n,\ell}^{(1)}\right)^2\rangle$ from this perspective. By looking at the term proportional to $\log^2u$ in (\ref{crossdiv}) we find
\begin{equation}
\left. \frac{1}{8} \sum_{m,n} c_{m,n}^{(2)}H_n^{(m)}(z,\zb) \right|_{div} =
\frac{1}{z\zb(\zb-z)} {\cal D}_4 \left( \frac{u^3 v^2 (1-u-v)}{16 (z-\zb)^6} \right) \log^2(1-\zb)
\end{equation}
where ${\cal D}_4 = D(D-2)\bar D(\bar D -2)$. We proceed exactly as before. First we project over the contribution for a given twist
\begin{equation}
\left. \frac{1}{8} \sum_{m} c_{m,n}^{(2)} \bar H^{(m)}(\zb)\right|_{div} = \frac{1}{2\pi i}\oint \frac{dz}{z^2} \frac{k_{1-h_n}(z)}{r_{h}}
 {\cal D}_4 \left( \frac{u^3 v^2 (1-u-v)}{16 (z-\zb)^6} \right) \log^2(1-\zb)
\end{equation}
This integral is straightforward to perform for any given integer $n$. For instance, for $n=0$ we obtain 
\begin{equation}
\left. \frac{1}{8} \sum_{m} c_{m,0}^{(2)} \bar H^{(m)}(\zb)\right|_{div} = 36\frac{6-6\zb+\zb^2}{\zb^2}\log^2(1-\zb)
\end{equation}
Plugging this into the formula (\ref{LSPTinversion}) and dividing by $a^{(0)}_{0,\ell}$ we obtain
\begin{equation}
\left(\gamma_{0,\ell}^{(1)}\right)^2 = \frac{576}{(\ell+1)^2(\ell+6)^2}
\end{equation}
where we have dropped the expectation value since there is no mixing for $n=0$.
This agrees with the correct value, for all values of the spin.
We can carry on this procedure for any desired value of $n$. The results are in perfect agreement with those of section \ref{inversion}. This is of course fully expected, since the divergence was computed from $ \avg{a^{(0)}\big(\gamma^{(1)}\big)^2}_{n,\ell}$, but it is a non-trivial test of our methods and it shows how they work. Next, let us turn to the combination containing $\gamma_{n,\ell}^{(2)}$, which is the main object of interest in this paper. This can be computed by substracting the first term in (\ref{H2dec}) to the total discontinuity, and then looking into the piece proportional to $\log u$. From here we repeat exactly the same steps as above. In order to make contact with section \ref{inversion} we focus in the combination that leads to $S_{n,\hb}^{(2)}$. In the language of the expansions above this is given by 
\begin{equation}
S_{n,\hb}^{(1)} = \sum_m\left( \frac{d_{m,n}^{(2)}}{J^{2m}} +\frac{1}{2}\frac{\partial_n c_{m,n}^{(2)} }{J^{2m}} \right)
\end{equation}
For the first few twists we obtain
\ba
S_{0,\hb}^{(1)}&=&\frac{96(16J^2-635)}{(J^2-6)(J^2-20)}\\
S_{1,\hb}^{(1)}&=&\frac{480 (-91710 - 2403 J^2 + 56 J^4)}{(-30 + J^2) (-12 + J^2) (-6 + J^2)}\\
S_{2,\hb}^{(1)}&=&\frac{1440 (-12910968 - 597906 J^2 - 6073 J^4 + 134 J^6)}{(-42 + J^2) (-20 + J^2) (-12 + J^2) (-6 + J^2)}
\ea
and so on. Again, the results are in perfect agreement with those obtained in section \ref{inversion}. Following the same procedure, we can compute the combination involving the OPE coefficients $a_{n,\ell}^{(2)}$. In this case the relevant combination is
\begin{equation}
S_{n,\hb}^{(0)} = \sum_m\left( \frac{e_{m,n}^{(2)}}{J^{2m}} +\frac{1}{2}\frac{\partial_n d_{m,n}^{(2)} }{J^{2m}} +\frac{1}{8}\frac{\partial^2_n c_{m,n}^{(2)} }{J^{2m}}\right)
\end{equation}
Our results are again in full agreement with those obtained in section \ref{inversion}. Since they are quite bulky, they are given in appendix \ref{app:ope}. Let us close this discussion with the following remark. We can rewrite the decomposition (\ref{H2dec}) as follows:
\begin{equation}
{\cal H}^{(2)}(z,\zb) = \sum_{m,n} s_{m,n}^{(0)}H_n^{(m)}(z,\zb)+\frac{1}{2} \sum_{m,n} s_{m,n}^{(1)}\partial_n H_n^{(m)}(z,\zb)+ \frac{1}{8} \sum_{m,n}  s_{m,n}^{(2)} \partial^2_nH_n^{(m)}(z,\zb) 
\end{equation}
Then the coefficients $s_{m,n}^{(p)}$ are the ones appearing in the large $J$ expansion of $S_{n,\hb}^{(p)}$. Note that given the factorisation properties of TCB taking the derivative w.r.t. to $n$ is a well defined operation and straightforward, since the $n$ dependence is explicit. This decomposition generalises to arbitrarily high orders in $1/c$.  

\subsection{From large spin perturbation theory to Froissart-Gribov inversion integral}

Using the method advocated in this section we can ask the following question. Which precise OPE data $a_{n,\ell}$ produces a given double discontinuity?
Our approach would be to express $a_{n,\ell} =a_{n}(J)$ as a series in inverse powers of $J^2$:
\begin{equation}
a_{n}(J) = \sum_m \frac{c_{m,n}}{J^{2m}}
\end{equation}
and use the technology developed in \cite{Alday:2016njk} to find all coefficients $c_{m,n}$. As we have seen, the problem can be factorised into the twist, or $n$, dependence and the dependence on the conformal spin. The first step is to project over a given twist. This can be done with the projectors introduced above and leads to
\begin{equation}
\label{spinLSPT}
\sum_m c_{m,n} \bar H_n^{(m)}(\zb) = \frac{1}{2\pi i}\oint \frac{dz}{z^2} \frac{k_{1-h_n}(z)}{r_{h_n}} z \zb(\zb-z) G(z,\zb)
\end{equation}
One can explicitly check that this procedure is equivalent to that of section \ref{inversion}. More precisely if we define the following function
\begin{equation}
c(h)= \int_0^1 \frac{dz}{z^2} \frac{k_{1-h}(z)}{r_{h}} g(z)
\end{equation}
then $c(h)$ has poles at $h=h_n$ for $n=0,1,\cdots$ and its residues are given by
\begin{equation}
{\rm Res}_{h=h_c} c(h)= \frac{1}{2\pi i}\oint \frac{dz}{z^2} \frac{k_{1-h_n}(z)}{r_{h_n}} g(z)
\end{equation}
In cases where $c(h)$ contains higher order poles, the subleading poles will correspond to derivatives of TCB w.r.t. $n$, and the parallel between the two methods is clear. 

The second step is to compute the dependence on the conformal spin. As shown in appendix \ref{apTCB} we can invert the problem (\ref{spinLSPT}). This results in an integral expression for the resumed $a_{n}(J)$. This involves the double integral (\ref{LSPTinversion}) and leads to
\begin{equation}
a_n(J)=4 \int_0^1 dt_1 \int_0^\infty dt_2 (1-t_1)^{\hb-1}(1+t_2)^{-\hb} \left. {\rm dDisc}\,\left[z \zb(\zb-z) \GG(z,\zb)\right] \right|_{\zb = \frac{1}{1+t_1 t_2}}
\end{equation}
where we leave implicit the projection over the twist $\tau_n$ and recall $J^2 =\hb (\hb-1)$. In order to make contact with the Froissart-Gribov inversion formula used in section \ref{inversion}, and prove the equivalence between the two methods, we consider the following change of variables:
\begin{equation}
t_1 = t\frac{1-\zb}{1-t \zb},~~~t_2 = \frac{1-t \zb}{t \zb}
\end{equation}
The integral over $t$ can be performed and we arrive to 
\begin{equation}
a_n(J)= 2 \int_0^1 d\zb  \zb^{\hb-2} \frac{r_{\hb}}{\hb-1/2} ~_2F_1(\hb,\hb;2 \hb;\zb)  {\rm dDisc}\,\left[ z \bar z(\zb-z) \GG(z,\zb)\right]
\end{equation}
where again, the projection over the twist is implicit. This exactly reproduces the analogous Froissart-Gribov inversion formula. 

\begin{figure}
\centering
 \hspace{-10mm}\begin{tabular}{c@{\hspace{-10mm}}c}
      \begin{minipage}{0.6\textwidth}
          \centering \hspace{2.4mm}\def\svgwidth{0.6\textwidth}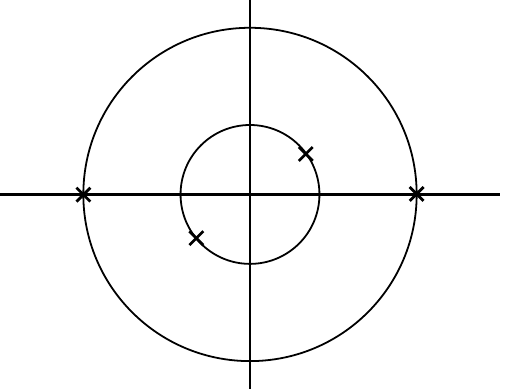\end{minipage}
          &
          \begin{minipage}{0.4\textwidth}
          \centering \includegraphics[width=0.45\textwidth]{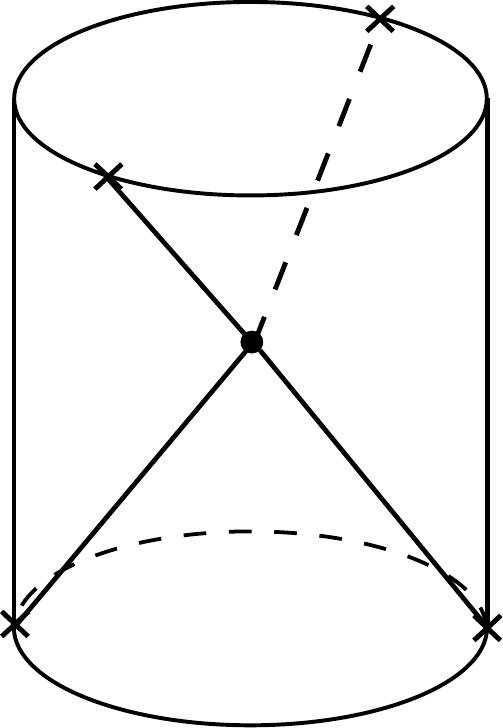}\end{minipage}\vspace{2mm}\\
          (a) & (b)
\end{tabular}   
\caption{(a) Four-point kinematics in the complex $\rho$-plane.
(b) The kinematics on the Lorentzian cylinder. In the ``bulk-point'' limit $z\to\zb$, particles
are effectively beamed onto a point in the AdS interior of the cylinder.}
\label{fig:rho}
\end{figure}

\section{Analytic results in the flat space limit}
\label{flat}

In this section we study the flat space limit of the one-loop correlator obtained so far,
and compare it to flat-space ten-dimensional supergravity.

Flat space physics can be accessed using suitable wavepackets focused onto a point in the bulk \cite{Heemskerk:2009pn,Gary:2009ae,
Maldacena:2015iua}.
To be self-contained, we give a brief exposition.
The kinematics are most conveniently described using the radial coordinates introduced in \cite{Hogervorst:2013sma}
and depicted in fig.~\ref{fig:rho}, see also section 5 of \cite{Maldacena:2015iua}.
In Euclidean kinematics, $\rho$ and $\rhob$ are complex conjugate to each other and
$\log (1/\rho\rhob)$ represents time in radial quantization. The bulk-point limit exists in real Minkowski signature
where that time is taken imaginary and approaches $i\pi$. Adding a rotation by $\pi$ to the scattering angle $\theta$, the corresponding
cross-ratios are
\be
  \rho = \frac{1-\sqrt{1-z}}{1+\sqrt{1-z}} =e^{i\theta-2\pi i+i x},\qquad
  \rhob=\frac{1-\sqrt{1-\zb}}{1+\sqrt{1-\zb}} = e^{-i\theta+ix}, \label{cross-ratios bulk point}
\ee
where $x\to 0$ in the limit (which implies $z,\zb\to \frac{2}{1+\cos\theta}$ with $z-\zb\sim x$).
Fast particles moving at the speed of light can then scatter in the bulk while conserving momentum.
Since the bulk point limit lies at the boundary of the $s$-channel OPE radius of convergence
($|\rho|\leq 1$, $|\rhob|\leq 1$ \cite{Hogervorst:2013sma}), singularities in this limit are tied
to the tail of the sum.

If one were to ignore the phase $e^{-2\pi i}$ representing the time evolution,
the correlator would admit the usual decomposition in super-conformal blocks
\begin{equation}
{\cal H}(u,v) = \sum_{\Delta,\ell} a_{\Delta,\ell} \,g_{\Delta,\ell}(u,v)\,. \label{bulk_lim_OPE0}
\end{equation}
The time evolution has the effect of multiplying each conformal block by a phase
\begin{equation}
{\cal H}_{cont.}(u,v) = \sum_{\Delta,\ell} e^{-i \pi(\Delta-\ell)}\, a_{\Delta,\ell} \,g_{\Delta,\ell}(u,v) \label{bulk_lim_OPE1}
\end{equation}
To understand the tail of the sum one may use the asymptotics of the hypergeometric functions (see \cite{Gary:2009ae}):
\be
 \lim_{h\to\infty} k_h(z) = \frac{(4\rho)^{h}}{\sqrt{1-\rho^2}} \left(1+ O(1/h)\right) \label{k_at_large_h}
\ee
which leads to the large-$n$ behavior (for any $x$)
\be
 \lim_{n\to\infty} a^{(0)}_{n,\ell} \,g_{n,\ell} = \frac{-64i\pi n^2}{z\zb(\zb-z)} \frac{(\ell+1)\sin((\ell+1)\theta)}{\sqrt{\sin^2\theta-\sin^2x}}\, e^{ix(2n+\ell+6)}.
\ee
One sees that each block has a $1/(\zb-z)\sim 1/x$ singularity.
However, a stronger singularity can be caused by the large-$n$ tail of the sum.
In a non-perturbative regime the extra phases in ${\cal H}_{cont.}(u,v)$ have been conjectured
to display a chaotic behaviour, ensuring that the singularity of the correlator is not enhanced compared with that
of individual blocks.  In a large $N$ perturbative regime this is not true anymore, since phases are small and in fact quite regular.
In the following we will focus on the dominant singularity at $x\to 0$ at each order in the $1/c$ expansion.
In this limit the dependence of the blocks on anomalous dimensions can be neglected as it produces subleading $d/dn$ terms,
and the above gives simply (see \cite{Heemskerk:2009pn})
\be
z\zb(\zb-z)\,\GG_{cont.}(u,v) \approx -64i\pi \sum_{n} n^2\,e^{2ixn} \sum_{\ell\,\text{even}}(\ell+1)^2 P_\ell(\cos\theta) 
\frac{\avg{ae^{-i\pi\gamma}}_{n,\ell}}{\avg{a^{(0)}}_{n,\ell}} \label{G_small_x}
\ee
where $P_\ell(\theta) = \frac{\sin (\ell+1)\theta}{(\ell+1)\sin \theta}$ are a four-dimensional version of Legendre polynomials.
This formula can be readily tested at the leading order: with the anomalous dimension $\gamma^{(1)}\approx \frac{-n^3}{2(\ell+1)}$
one finds $z\zb(\zb-z)\,\GG^{(1)}_{cont.} \approx \frac{-30\pi^2}{x^6\sin^2\!\theta}$, which is in precise agreement with the analytic
continuation of the $\bar{D}$ function in eq.~(\ref{Gtree}).

\subsection{Large-$n$ limit of CFT-data}
This discussion motivates to study the averages $\avg{ae^{-i\pi\gamma}}_{n,\ell}$ in the large $n$ limit. We will do so in two different ways. First from our explicit results, and then directly from the inversion integral. 

\subsubsection{Large-$n$ limit from explicit results}
Up to order $1/c^2$ the average in question is equivalent to 
\begin{eqnarray}
\avg{ae^{-i\pi\gamma}}_{n,\ell} &=& \avg{a^{(0)}}_{n,\ell} + \frac{1}{c} \left( \avg{a^{(1)}}_{n,\ell} - i \pi \avg{a^{(0)} \gamma^{(1)}}_{n,\ell}   \right) \\ 
& & + \frac{1}{c^2}\left( \avg{a^{(2)}}_{n,\ell} -i \pi   \avg{a^{(1)}\gamma^{(1)} + a^{(0)}\gamma^{(2)}  }_{n,\ell}  -\frac{\pi^2}{2}  \avg{a^{(0)}\big(\gamma^{(1)} \big)^2}_{n,\ell}  \right) \nonumber
\end{eqnarray}
The last term to order $1/c^2$ is simply $S_{n,\hb}^{(2)}$. Its large $n$ behaviour is given by
\begin{eqnarray}
S_{n,\hb}^{(2)} &=&\frac{1}{180} \left(2 \ell^4+8 \ell^3+22 \ell^2+28 \ell+15\right)\left(\psi ^{(0)}\left(\frac{\ell}{2}+1\right) -\psi ^{(0)}\left(\frac{\ell}{2}+\frac{1}{2}\right)\right) n^{12} \nonumber \\
& & - \frac{4 \ell^4+18 \ell^3+50 \ell^2+71 \ell+41}{360 (\ell+1)} n^{12}  + \cdots 
\end{eqnarray}
It is remarkable that although $ \gamma_{n,\ell}^{(1)} \sim n^3$ the average of its square grows like $n^{11}$. This is only possible thanks to mixing and as we will see is necessary for matching with quantum gravity in flat space.  Roughly speaking the additional power of $n^5$ stems from the S${}_5$ factor behaving
like five additional flat dimensions in that limit. Next we recall the relation between the second average and $S_{n,\hb}^{(1)}, S_{n,\hb}^{(2)}$ studied in sections \ref{inversion} and \ref{LSPT}:
\begin{equation}
\avg{a^{(0)} \gamma^{(2)}+ a^{(1)} \gamma^{(1)}}_{n,\ell} = S_{n,\hb}^{(1)} +\frac{1}{2} \partial{}_{\hb} S_{n,\hb}^{(2)}
\end{equation}
The large $n$ behaviour of $S_{n,\hb}^{(1)}$ is more complicated. For a given $n$ the result can be written as a sum over poles in $J^2$:
\begin{equation}
S_{n,\hb}^{(1)} = \sum_{p=-2}^{n-2} \frac{\kappa_n^2 r(n,p)}{J^2-(n-p)(n-p+1)} + \frac{\kappa_n^2 d_n}{J^2-(n+4)(n+5)}
\end{equation}
where recall $J^2=(n+\ell+3)(n+\ell+4), \kappa_n=(n+1)(n+2)(n+3)(n+4)$ and
\begin{equation}
d_n=-\frac{\left(n^2+7 n+12\right) \left(n^2+11 n+30\right)^2}{576 (2 n+5)}
\end{equation} 
The form of the residues $r(n,p)$ is much more complicated. For any fixed $p$ they can be expressed in terms of polygamma functions and polynomials whose degree increase with $p$. In principle one could use eq.~(\ref{S1_alln}), but
alternatively we have found that $r(n,p)$ satisfies a complicated recursion relation, relating $r(n,p)$ to $r(n-2,p-2)$. This recursion relation has the following structure
\begin{equation}
P^{(8)}(p,n) \kappa_n r(n,p)+ \tilde P^{(8)}(p,n)  \kappa_{n-2} r(n-2,p-2) + R(p,n) =0
\end{equation}
where $P^{(8)}(p,n), \tilde P^{(8)}(p,n) $ are polynomials in $p,n$ of total degree 8, and $R(p,n)$ is a rational functions which depends on whether $p$ is even or odd. This recursion relation allows to write an arbitrary number of residues once the first ones are known, and hence they allow to reconstruct the full $S_n$. Before proceeding, note that the expression can also be written as a sum over poles on $\ell$. Indeed 
\begin{equation}
J^2-(n-p)(n-p+1)=(3+p+\ell)(4+2n-p+\ell)
\end{equation}
We would like to use the above recursion relations to compute the behaviour of $S_n$ for large $n$ and finite spin. This computation is not straightfoward. One of the reasons is that all poles contribute in the large $n$ limit and there are subtle cancelations between $p$ odd and even. In order to proceed, we will consider the following transform
\begin{equation}
S_{n,\hb}^{(1)}= \int_0^1d\zeta \zeta^{\ell-1} f_n(\zeta)
\end{equation}
we have traded the spin dependence by the dependence on $\zeta$. Single poles on $\ell$ map to positive powers of $\zeta$
\begin{equation}
\frac{1}{3+p+\ell}= \int_0^1d\zeta \zeta^{\ell-1} \zeta^{p+3}
\end{equation}
For a fixed $n$, $f_n(\zeta)$ is a polynomial of degree $2n+6$. Reciprocity for $S_{n,\hb}^{(1)}$ implies the following symmetry 
\begin{equation}
f_n(\zeta)=-\zeta^{2n+7}f_n(1/\zeta)
\end{equation}
One of the reasons to use this representation is that as $n$ increases, $f_n(\zeta)$ is much better behaved that the (infinite) sum over poles. We would like to study this function in the large $n$ limit, in the range $0<\zeta<1$. The recursion relations above imply the following expansion
\begin{equation}
f_n(\zeta)=n^{12} f^{(0)}(\zeta) + \cdots
\end{equation}
and we would like to find $f^{(0)}(\zeta)$. In order to do so, we study the recursion relations for fixed $p$ in the large $n$ limit. This allows to find the coefficient $c_p$ in front of $\zeta^{p+3}$ in the small $\zeta$ expansion of $f^{(0)}(\zeta)$. For instance, for odd $p$ we find
\begin{eqnarray*}
c_p =& &\frac{1}{90} \left(2 p^4+16 p^3+58 p^2+104 p+75\right)\left( \psi ^{(0)}\left(\frac{p}{2}+1\right) -\psi ^{(0)}\left(\frac{p+3}{2}\right) \right)\\
& & -\frac{12 p^4+94 p^3+296 p^2+431 p+232}{180 (p+2)}
\end{eqnarray*}
The expression for $p$ even is a bit more complicated. Having found $c_p$ we can perform the sum, which leads to the final expression for $f^{(0)}(\zeta)$
\begin{eqnarray*}
f^{(0)}(\zeta) &= &\frac{(\zeta-1)^5 (\zeta+1)^5}{61440 \zeta^4}\log(1+\zeta)-\frac{(\zeta-1)^5 \zeta}{60 (\zeta+1)^5}\log 2\\
& &-\frac{(\zeta-1)^7 \left(\zeta^8+12 \zeta^7+68 \zeta^6+244 \zeta^5+630 \zeta^4+244 \zeta^3+68 \zeta^2+12 \zeta+1\right)}{61440 \zeta^4 (\zeta+1)^5}\log(1-\zeta)\\
& &-\frac{(\zeta-1)^5 \left(\zeta^2-1\right)^2 (\zeta (\zeta (\zeta (\zeta+10)+34)+10)+1)}{30720 \zeta^3 (\zeta+1)^5}
\end{eqnarray*}
This gives the full spin dependence of the leading term $S_{n,\hb}^{(1)} \sim n^{12}$. Combining this with the result for $S_{n,\hb}^{(2)}$ give us the full spin dependence of the average $\avg{a^{(0)} \gamma^{(2)}+ a^{(1)} \gamma^{(1)}}_{n,\ell}$. Let us quote the results for a few values of the spin
\begin{eqnarray}
\avg{a^{(0)} \gamma^{(2)}+ a^{(1)} \gamma^{(1)}}_{n,2}  &=& \left( \frac{17411 \pi ^2}{147456}-\frac{4189}{8640}+\frac{17 \log ^2 2}{12}-\frac{707 \log 2}{360} \right) n^{12}+ \cdots \\
\avg{a^{(0)} \gamma^{(2)}+ a^{(1)} \gamma^{(1)}}_{n,4}  &=& \left(\frac{171007 \pi ^2}{245760}-\frac{719657}{252000}+\frac{167 \log ^2 2}{20}-\frac{5209 \log 2}{450} \right) n^{12}+ \cdots \nonumber
\end{eqnarray}
Finally, it turns out the average $\avg{a^{(2)}}_{n,\ell}$ is subleading for large $n$, and will not be important for our purposes. 
This is expected since the Euclidean correlator (\ref{bulk_lim_OPE0}) has no $x\to 0$ singularity at generic angle
(and only a mild one at zero angle).

\subsubsection{Large-$n$ limit from inversion integral}
\label{ssec:large n inversion}

The simplicity of the preceding result suggests a more direct route and in fact we now
show how to take this limit directly from the Froissart-Gribov inversion integral (\ref{fg}).
The key fact is that the poles in $c(h,\hb)$ originate only from the $z\to 0$ limit of integration.
Therefore just by rotating the $z$ contour clockwise by $2\pi$, and dropping an arc at $|z|=1$ which produces no pole,
we can eliminate the phase:
\ba
e^{-2\pi ih}\cc{h}{\hb} &=& \int_0^1 \frac{dz}{z^2} \frac{k_{1-h}(z)}{r_{h}}\,
 \int_0^1 \frac{d\zb}{\zb^2}  \frac{r_\hb}{\hb-\frac12}k_{\hb}(\zb) \frac{{\rm dDisc}\,[z\zb(\zb-z)\GG(z^\circlearrowright,\zb)]}{4\pi^2} + \mbox{pole-free}
 \nl &\equiv& c^\circlearrowright(h,\hb)\,.
\label{ccirc}
\ea
The notation indicates that the correlator is evaluated with $z$ rotated clockwise around the origin.
Recall that the double-discontinuity  (\ref{ddisc}) is itself computed as an analytic continuation, but with respect to the other variable (around $\zb=1$),
so these two continuations commute with each other.

Our interest is in the asymptotic spectral density of $\cc{h}{\hb}^\circlearrowright$.
This can be defined mathematically by taking the difference slightly above and below the real axis
\be
 \avg{a e^{-i\pi\gamma}}_{n,\ell} \approx \frac{1}{i \pi} \left( c^\circlearrowright(h\times e^{i\alpha},\hb)
- c^\circlearrowright(h\times e^{-i\alpha},\hb)\right)
\ee
where $\alpha>0$ is a small phase.
This analytic function is what would enter, for example, in the Watson-Sommerfeld representation in appendix B to \cite{Heemskerk:2009pn}.
(The spurious poles mentioned below eq.~(\ref{z_int}) can be neglected in the limit.)
For the first term, one sees that the integral (\ref{ccirc}) would decay exponentially if the $z$ contour could be rotated clockwise,
however this is obstructed by the singularity at $z=\zb$.  The second term however decays exponentially because there are no singularities
obstructing a rotation in the other direction.  The conclusion is that the large-$n$ behavior of the coefficients is controlled by the
singularity closest to the origin, in this case $z=\zb$.
Near this point, setting $z=\zb+2x\zb\sqrt{1-\zb}$ with $x\to 0$, we find that the double-discontinuity
(\ref{dDisc2}) diverges like
\be
 \frac{{\rm dDisc}\,\left[z\zb(\zb-z)\GG^{(2)}(z^\circlearrowright,\zb)\right]}{4\pi^2} \to 2\pi i \times \frac{13!}{(2x)^{14}} \times g_2(\zb),
\ee
with
\be
\label{CFTdisc}
g_2(z)= \frac{z(1-z)^2}{480}
\left(\left(1-\frac{1}{z^5}\right)\log(1-z) + \frac{1}{z} -\frac{1}{z^4}+\frac{1}{2z^2}-\frac{1}{2z^3} +i\pi -\log(z)\right)\,.
\ee
Using the asymptotic formula (\ref{k_at_large_h}), the integral (\ref{ccirc}) thus gives
\be
\frac{\avg{ae^{-i\pi\gamma}}_{n,\ell}^{(2)}}{\avg{a^{(0)}}_{n,\ell}}
\to \frac{1}{n^2(\ell+1)} \int_C \frac{dx}{2\pi i}\frac{13!}{(2x)^{14}} e^{-2nx}
\int_0^1 \frac{d\zb}{\zb^2} \left(\frac{1-\sqrt{1-\zb}}{1+\sqrt{1-\zb}}\right)^{\ell+1} g_2(\zb)
\ee
where the contour $C$ is a ``keyhole'' contour encircling the origin to the right clockwise.
We see that the inversion integral is neatly factorized,
in perfect parallel with the OPE (\ref{G_small_x}): the variable $\zb$ ranges between 0 and 1
and provides the angular dependence, while the distance $x$ to the singularity is conjugate to $n$.
The integral over $x$ produces a power of $n$, and we thus find the following high-energy behavior:
\be
\frac{\avg{ae^{-i\pi\gamma}}_{n,\ell}}{\avg{a^{(0)}}_{n,\ell}}
\xrightarrow{n\to\infty} 1 + \frac{1}{c} \frac{i\pi \,n^3}{2(\ell+1)} +
\frac{1}{c^2}\frac{i\pi\,n^{11}}{\ell+1}\int_0^1 \frac{g_2(\zb)\,d\zb}{\zb^2} \left(\frac{1-\sqrt{1-\zb}}{1+\sqrt{1-\zb}}\right)^{\ell+1}
+ O(1/c^3)\,. \label{asympt_a}
\ee
This can be integrated analytically in terms of harmonic sums but we did not find the result particularly illuminating.
For the first few values of the spin the average at order $1/c^2$ gives
\begin{eqnarray*}
\avg{ae^{-i\pi\gamma}}_{n,2} &=&i\pi \left(\frac{4189}{8640}-\frac{17411 \pi ^2}{147456}-\frac{17}{12}\log 2\big(\!\log 2-i\pi\big)+\frac{707}{720}\big(2\log 2-i\pi\big)\right) n^{12}+ \cdots \\
\avg{ae^{-i\pi\gamma}}_{n,4} &=&i\pi \left(\frac{719657}{252000} -\frac{171007 \pi ^2}{245760}-\frac{167}{20}\log 2\big(\!\log 2-i\pi\big)+\frac{5209}{900}\big(2\log 2-i\pi\big)\right) n^{12}+ \cdots,
\end{eqnarray*} 
in perfect agreement with our previous results. Given that these two calculation methods were both subtle in different ways,
it is nontrivial and very reassuring that they agree!  We shall now compare
this result with flat space supergravity. The latter representation (\ref{asympt_a}) will turn out particularly convenient for that scope.

\subsection{Comparison with flat space supergravity amplitude}
\label{analytic}

The relation between CFT correlators and the S-matrix of the higher-dimensional bulk theory
has been analyzed in many works, see {\it e.g.} \cite{Gary:2009ae,Heemskerk:2009pn,Penedones:2010ue,Okuda:2010ym,Maldacena:2015iua}.
The idea is that in the $x\to 0$ limit described above, the CFT correlator 
effectively focuses particles at each other in the bulk.
By analyzing bulk Landau diagrams, a precise relation between the $z\to\zb$ singularity and the high-energy
behavior of the amplitude is obtained.
Setting $\Delta=2$ and $d=4$ in eq.~(5.5) of \cite{Maldacena:2015iua} (see also \cite{Gary:2009ae})
this relation reads:
\be
z\zb(\zb-z)\,\GG_{\rm cont} = \frac{1}{2} \int_0^\infty \omega^2d\omega \,e^{2i\omega x}\, \sqrt{s}A_5(s,t)\,,
\label{G_from_flat_space}
\ee
where $\omega$ represents the energies of each incoming particle in units of the AdS radius $L$,
the Mandelstam variable $s=4\omega^2/L^2$ is the center-of-mass energy, and $-t/s=\frac{1-\cos\theta}{2}$ encodes the scattering angle.
This formula gives the leading singular term at $x\sim z-\zb\to 0$ for each order in $1/c$.
To make contact with the OPE on the CFT side,  we use the usual partial wave expansion
for the five-dimensional (flat space) amplitude $A_5(s,t)$:
\be
 iA_5(s,t) =  \frac{128\pi}{\sqrt{s}} \sum_{\ell\, \text{even}} (\ell+1)^2 b_\ell(s)\, P_\ell(\cos\theta) 
 \label{partial_waves_flat}
\ee
with $\cos\theta=1+\frac{2t}{s}$ and $P_\ell$ is as defined below eq.~(\ref{G_small_x}). The prefactor is simply one over the
phase space volume for two identical particles, ensuring that $b_\ell^{(0)}=1$ in the absence of interactions.
Comparing (\ref{G_small_x}), (\ref{G_from_flat_space}) and (\ref{partial_waves_flat}) gives a key formula,
quoted in section \ref{generalities}, to compare the OPE data and flat space amplitude:
\be
 \lim_{n\to\infty} \frac{\avg{a e^{-i\pi\gamma}}_{n,\ell}}{\avg{a^{(0)}}_{n,\ell}} = b_\ell(s)\,, \qquad L\sqrt{s}=2n\,.
 \label{flat_space_matching}
\ee

\begin{figure}
 \centering
 \def\svgwidth{0.6\textwidth}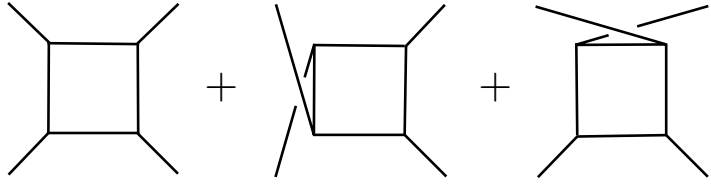
\caption{The one-loop amplitude in ten-dimensional type IIB supergravity is the sum of three scalar box integrals.
}
\label{fig:boxes}
\end{figure}

At energy scales between the AdS and string scales, the flat space amplitude $A_5(s,t)$
can be reliably computed using perturbative quantum gravity in flat space, viewed as an effective field theory.
We notice that in this regime the AdS${}_5\times$S${}_5$ geometry is fundamentally ten-dimensional,
so the relevant effective theory is ten-dimensional IIB supergravity.
Fortunately, the one-loop flat space integrand in this theory, incorporating graviton and gravitino loops,
was worked out long ago \cite{Green:1982sw,Bern:1998ug}.
It is a simple sum of scalar boxes (see fig.~\ref{fig:boxes}), thanks to the so-called no-triangle property of maximal supergravity:
\be \label{A10}
 A_{10}^{\text{sugra}}(s,t) = 8\pi G_N \frac{s^3}{t u} + \frac{(8\pi G_N)^2}{(4\pi)^5}\left(I_{box}(s,t)+I_{box}(s,u)+I_{box}(t,u)\right) +O(G_N^3)\,,
\ee
where $G_N=\frac{\pi^4 L^8}{8c}$
is the ten-dimensional Planck constant, with $c=\frac{N^2-1}{4}$ and $L$ the AdS radius.
To be fully precise, let us specify which polarization we have chosen for the external gravitons:
to match with the correlator $\GG^{(105)}$, which corresponds to two identical complex scalars,
one should choose the polarizations of gravitons $1$ and $2$ to be two identical null tensors orthogonal to all the momenta.
(By supersymmetry, all other choices are equivalent up to an overall factor.)

The box integral has quadratic and logarithmic divergences but can be readily evaluated using e.g. dimensional regularization.
The logarithmic divergence nicely cancels in the sum over boxes due to the relation $s+t+u=0$,
so the only ambiguity is the quadratic divergence (which has to be restored manually in dimensional regularization), which we'll interpret shortly.
Note that even though the integral is computed in ten dimensions, we need to expand the result in five-dimensional partial waves in order to compare with the CFT${}_4$ data
in eq.~(\ref{flat_space_matching}).  We obtain $A_5$ simply by dividing by the volume of the sphere, vol(S${}_5)=\pi^3L^5$.
Using the result for the box integral recorded in eq.~(\ref{Ibox}), we thus obtain
\be
 \frac{\sqrt{s}}{64\pi^2}A_{5}^{\text{sugra}}(s,\cos\theta) = \frac{(\sqrt{s}L/2)^3}{c}  \frac{1}{2\sin^2\theta}
 + \frac{(\sqrt{s}L/2)^{11}}{c^2} f_2\left(\frac{1+\cos\theta}{2}\right) +O(1/c^3) \label{A5_sugra}
\ee
where $f_2(x)$ contains the angular dependence and is given in eq.~(\ref{f2}).
From this, the partial wave coefficients are obtained simply
by inverting (\ref{flat_space_matching}) using the orthogonality of the polynomials $P_\ell(\cos\theta)$:
\be
 b_\ell(s) = 1 + \frac{i\pi}{\ell+1} \int_0^\pi \frac{d\theta}{\pi}\, \sin\theta\sin((\ell+1)\theta) \,\frac{\sqrt{s}}{64\pi^2}A_5^{\text{sugra}}(s,\cos\theta). \label{bl_inverse}
\ee
An excellent way to perform such integrals is to use the Froissart-Gribov method and deform the contour in $x=\frac{1+\cos\theta}{2}$
so that it picks the singularities of $A_5$.
Physically, this method of reconstructing partial waves from cuts is equivalent to a dispersion relation, as discussed for example in section 2.5 of \cite{Caron-Huot:2017vep}.
The $t$- and $u$-channel branch cuts at $x>1$ and $x<0$ give the same by symmetry, and we get
\be
 b_\ell(s) = 1+ \frac{1}{c} \frac{i\pi\,(\sqrt{s}L/2)^3}{2(\ell+1)} +
 \frac{1}{c^2}\frac{i\pi\,(\sqrt{s}L/2)^{11}}{\ell+1}
 \int_0^1 \frac{d\zb}{\zb^2} \left(\frac{1-\sqrt{1-\zb}}{1+\sqrt{1-\zb}}\right)^{\ell+1} {\rm Disc}_t f_2(1/\zb), \label{bl_fg}
\ee
where $\zb=1/x$ and the $t$-channel discontinuity is the difference between going above or below the $t$-channel cut at $x>1$:
\be\begin{aligned}
 {\rm Disc}_t f_2(1/z) &\equiv \frac{2}{i\pi}\left( f^{(2)}(1/z+i0)-f^{(2)}(1/z-i0)\right) \\
&= \frac{z(1-z)^2}{480}\left(\left(1-\frac{1}{z^5}\right)\log(1-z) + \frac{1}{z} -\frac{1}{z^4}+\frac{1}{2z^2}-\frac{1}{2z^3} +i\pi -\log(z)\right)\,. \label{disc f2}
\end{aligned}\ee
We have verified the agreement between (\ref{bl_inverse}) and (\ref{bl_fg}) to high numerical accuracy for a variety of spins $\ell$.

\begin{figure}
 \centering\hspace{15mm}
 \def\svgwidth{0.8\textwidth}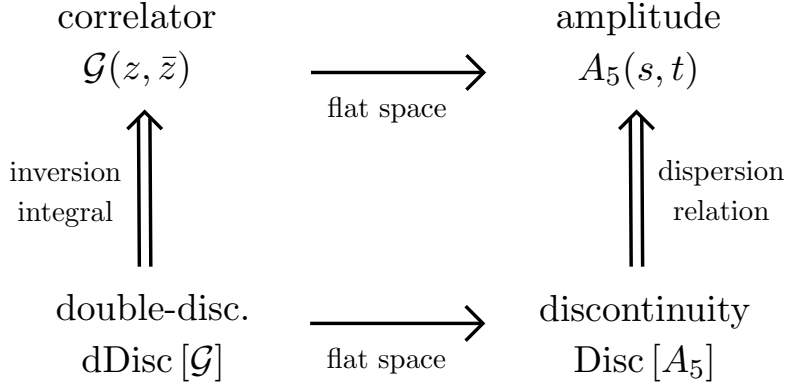 \\[5mm]
\caption{A commutative diagram which explains the agreement between the  one-loop CFT and supergravity calculations:
the discontinuities, which determine the outcome of both calculations, match each other.}
\label{fig:commutative}
\end{figure}

The discontinuity of the amplitude (\ref{disc f2}) is identical to the double-discontinuity of the correlator (\ref{CFTdisc}) obtained directly from the CFT!
In other words, the operations of taking discontinuities commute with taking the flat space limit, leading to the diagram shown in figure \ref{fig:commutative}.
Since the full answers are reconstructed from the singularities, the full answers also agree.
Comparing eqs.~(\ref{asympt_a}) and (\ref{bl_fg}),
we have matched not only the answers on both sides, but also the computation techniques.
This makes transparent the relation between our computation in CFT and the standard reconstruction of S-matrices via dispersion relations. 


\subsection{Local counter-terms and uniqueness of the reconstruction}
\label{uniqueness}

As noted in section \ref{inversion}, reconstructions via dispersion relations typically suffer from polynomial ambiguities,
which are supported on finitely many spins. These represent bulk contact interactions.  Such ambiguities potentially affect both
our tree and one-loop results, so let us address them quantitatively.
They were classified earlier in \cite{Alday:2014tsa} using Mellin space technique.
The simplest one is supported on spin 0.  In terms of the analytic function $c(h,h+\ell+1)$ defined in eq.~(\ref{residues}),
it can be directly constructed as follows.
First, single-valuedness (from an argument similar to \ref{ssec:large n inversion})
requires it to vanish like $e^{-2\pi |{\rm Im}\,h|}$  at large imaginary $h$.
Since it can have only at most double poles at the double-trace locations, we write
\be
 \delta c(h,h+1) = \frac{\pi^2}{\sin(\pi h)^2}\delta c'(n) \label{delta c}
\ee
where $\delta c'(n)$ is a rational function of $n=h-3$.
Second, shadow symmetry $\Delta\to 4-\Delta$ forces
$\delta c'(n)$ to be even under $n\to -6{-}n$, and it needs double zeros
at $n=-1,-2,-3,-4,-5$ in order to cancel the poles below the double-trace threshold.
The case $n=-3$ requires an extra zero to cancel a pole of the block (\ref{normalization}); poles at $n=-5/2$ and its reflection are exceptionally allowed, due to a zero of the block.
Thus, up to an overall factor, the minimal ambiguity is:
\be
\frac{\avg{a^{(0)}\delta \gamma}_{n,0}}{\avg{a^{(0)}}_{n,0}}
= -\frac{\delta c'(n)}{n+3}
 = -c_1 \frac{(n+1)^2(n+2)^2(n+3)^3(n+4)^2(n+5)^2}{960 (n+5/2)(n+7/2)}\,, \label{gamma c1}
\ee
whereas the residue of eq.~(\ref{delta c}) gives the derivative relation
$\avg{\delta a}_{n,0}=\frac12\partial_n \avg{a^{(0)}\delta \gamma}_{n,0}$.
Summing up the OPE, we then find that the corresponding ambiguity in the correlator,
\be
 \delta \mathcal{G}(u,v) = c_1 u^2 \bar{D}_{4,4,4,4}(u,v),
\ee
is indeed crossing-symmetric.  It is possible to show that any non-minimal solution, obtained by multiplying
the above by a polynomial, would require higher spins to satisfy crossing.

We would like to explain why the coefficient $c_1$ is small, from the CFT perspective.
We need to use supersymmetry, which relates the correlator
in the $105$ representation of SU(4)${}_{\cal R}$, discussed so far, to that in the singlet representation.
This relation has a nontrivial prefactor, which shifts the spin by 4 in some of the blocks, see eq.~(8.1) and the third line of eq.~(8.7) of \cite{Dolan:2001tt}.
This shift is important because for spins two and higher the Froissart-Gribov integral is guaranteed to converge to the correct CFT-data in any complete, unitary theory \cite{Caron-Huot:2017vep}.
Since this could be applied to the singlet correlator, we conclude that the \emph{full} CFT-data in $\mathcal{N}=4$ SYM is reconstructed from the discontinuities.
Furthermore, the nonperturbative bound in eq.~(5.4) there, applied to the singlet correlator,
constrains the contributions of heavy operators, neglected so far:
\be
 \Big| \delta c^{\rm singlet}_{h,h+\ell+1}\Big| \lsim \frac{1}{c} \frac{1}{(\Delta_{\rm gap}^2)^{\ell-2}}
 \quad \mbox{for $\ell\geq 2$}
 \quad\Rightarrow\quad
\Big| \delta c_{h,h+\ell+1}\Big| \lsim \frac{1}{c} \frac{1}{(\Delta_{\rm gap}^2)^{\ell+2}}
\quad \mbox{for $\ell\geq 0$}\,. \label{bound}
\ee
This implies for example $c_1\lsim \frac{1}{c\,\Delta_{\rm gap}^4} \sim \frac1c \frac1\lambda$ where $\lambda$ is the `t Hooft coupling.
This establishes uniqueness of the tree-level solution given above in the strong coupling limit $\lambda\to\infty$.
Note also that $c_1$ is strictly positive, since it is controlled by the integral of a positive-definite double-discontinuity. This conclusion was also reached in \cite{Alday:2016htq}. 

This bound on $c_1$ is essentially what one would have obtained from a naive application of the bound on chaos \cite{Maldacena:2015waa}
to the singlet correlator, see also \cite{Costa:2017twz}. The inversion integral however
naturally builds in the fact that the bound applies not to the CFT-data itself,
but to the \emph{difference} $\delta c$ between
the CFT-data and the analytic contribution to it from light $t$-channel operators.
The latter gave the supergravity result and is not small.

According to eq.~(\ref{flat_space_matching}), this small $\sim n^9$ ambiguity reduces, in the flat space limit,
to an 8-derivative contact interaction $\delta A_{5}^{\text{sugra}}(s) = c_1\frac{\pi^2s^4L^9}{3840}$,
corresponding to an $R^4$ contact interaction in gravity.
The fact that the simplest contact interaction has 8 derivatives follows simply from supersymmetry,
since the four-particle amplitude is proportional to a 16-dimensional $\delta$-function $\delta^{16}(Q)$, which evaluates to $s^4$
for the graviton polarizations described above.
In the CFT calculation, the role of this $\delta$-function was effectively played by the relation between SU(4)${}_{\cal R}$ representations.
This explains also why there is no logarithmic divergence at one-loop, since any crossing-symmetric 10-derivative interaction would be proportional to $s{+}t{+}u=0$.
However we can see
that the above bound is not optimal: the $R^4$ operator is known to appear with coefficient
$\frac1c\frac{1}{(\alpha')^3}\sim \frac1c\frac1{\lambda^{3/2}} \ll \frac1c \frac1\lambda$
in tree-level string theory (stringy corrections to CFT-data have been studied in \cite{Goncalves:2014ffa}).
The non-optimality grows for higher-derivative contact interactions, since the bound
controls the spin of the interactions rather than their mass dimensions.

At one-loop, a quadratically divergent $R^4$ interaction is forced on us, even at large $\lambda$, because its coefficient is given by a positive-definite
sum rule which diverges in the one-loop approximation.
Indeed the double-discontinuity vanishes at tree-level away from $\zb=1$ and starts at one-loop,
where we found a positive-definite result that for $\zb\sim z$ grows like
\be
{\rm dDisc}\,[\GG] \sim \frac{1}{c^2} \frac{1}{z^4}\,.
\ee
This growth can stop at $z\sim 1/\Delta_{\rm gap}^2$, where operators neglected so far, with $\Delta>\Delta_{\rm gap}$, enter the $t$-channel OPE.
Integrating the Froissart-Gribov formula over the region under control gives a contribution to OPE data
\be
c(h,h+\ell+1)\Big|_{\rm controlled}\propto \int_{\sim 1/\Delta_{\rm gap}^{2}}^1 dz z^{\ell+2} {\rm dDisc}\,[\GG] \sim \frac{1}{c^2}\frac{1}{\Delta_{\text{gap}}^{2(\ell-1)}}\,,
\ee
which diverges quadratically for $\ell=0$ (but converges for $\ell>1$). By crossing, this divergent part has to be proportional to the ambiguity (\ref{gamma c1}).
But since it comes from a positive-definite sum rule, it cannot be canceled by any counter-term in a complete, unitary theory,
even if we try to take the gap to infinity.  In fact, the gap can't be taken bigger than $c^{1/4}\sim m_{pl}^2L^2$,
in order for ${\rm dDisc}\,[\GG]$ to remain locally between 0 and 1, as it must.

In this extreme case where the growth continues up to ${\rm dDisc}[\GG]\approx 1$, nonperturbative contributions
to the CFT-data can have size at most $1/c^{\frac{3+\ell}{2}}$.
This corresponds to a scenario where the dual theory lacks any separation between the string and Planck scales.
There is still a good effective theory description at the AdS scale thanks to large-$N$, but higher-dimensional operators
in the effective Lagrangian assume values which can only be determined nonpertubatively (for the first operator, using dimensional analysis with Planck-scale suppression
one would expect $c_1\sim 1/c^{7/4}$, consistent with the above bounds).
It would be interesting to further study such theories of strongly coupled gravity, perhaps
by combining the present methods with the numerical bootstrap following \cite{Beem:2016wfs}.

\section{Conclusions}

In the present paper we have analysed the double-discontinuities of the four-point correlator of the stress-tensor multiplet in ${\cal N}=4$ SYM at large t'Hooft coupling and at order $1/c^2$, with $c \sim N^2$. We have explicitly shown how to extract the full CFT-data from these discontinuities. This was done by two alternative inversion procedures, one based on a Froissart-Gribov type inversion integral for CFT, and the other based on large spin perturbation theory. The procedures were then shown to be equivalent. Our computation makes explicit the fact that double-discontinuities contain all the relevant physical information. In particular they also allow to reconstruct the full correlator, without any additional assumptions about the space of functions that can appear. This is reminiscent of the  Kramers-Kronig relations that allow to reconstruct a holomorphic function that decays at infinity from its imaginary part. 

Via the $AdS/CFT$ duality $1/c^2$ corrections describe one-loop effects in the dual gravitational theory. We have considered the limit in which the CFT correlator is expected to reproduce the one-loop scattering amplitude of gravitons in ten-dimensional
supergravity in flat space, finding perfect agreement. An elegant way to summarise this agreement is that the flat space limit of the discontinuity of the CFT correlator agrees with the discontinuity of the flat space scattering. These discontinuities then determine both results: on the CFT side, through the inversion procedure presented here, and on the amplitude side through dispersion relations.  The 10-dimensional nature of the underlying string
theory is realized through the mixing between CFT operators with different SU(4)${}_{\cal R}$ charge.

The inversion presented in this paper makes clear which ambiguities may arise and how. In any non-perturbative theory it has been proven in \cite{Caron-Huot:2017vep}  that proper Regge behaviour constraints these ambiguities to spin lower than two, and we have shown that supersymmetry removes all such ambiguities.
In a perturbative expansion however the situation can be a bit worse but we have shown that there are no ambiguities at order $1/c$ in the limit of large `t Hooft coupling,
and only a single ambiguity for spin zero at order $1/c^2$.  This ambiguity corresponds to an $R^4$ counter-term, which is indeed
generated at one-loop in ten-dimensional supergravity with a quadratically divergent coefficient. Interestingly,
we find that fully cancelling this term by a negative counter-term would be inconsistent with unitary, hence
a large positive value must be assumed nonperturbatively.

There are several open problems that would be interesting to attack. First of all, our methods are completely general and one should be able to study amplitudes of more generic gravitational theories. It would be interesting to study theories in different number of dimensions and with less super-symmetry, and more broadly to see if this method
can be used to constrain higher-dimensional supergravity theories.  The interplay with new techniques for direct computations in AdS, see \cite{Cardona:2017tsw,Yuan:2017vgp}, should be pursued, as well as the prospect of fixing nonperturbatively the coefficients of higher-dimensional contact interactions.
It would also be interesting to study higher order corrections in $1/c$.

\acknowledgments{We would like to specially thank A. Bissi for collaboration at early stages of this work and in related topics. We would also like to thank Ofer Aharony for enlightening discussions.
We would like to thank the organisers of the Strings 2017 conference in Tel Aviv were this work was initiated.
The work of LFA was supported by ERC STG grant 306260. LFA is a Wolfson Royal Society Research Merit Award holder. 
SCH's work is supported by the National Science and Engineering Council of Canada.
}

\appendix

\section{The double-discontinuity of the one-loop correlator}
\label{app:ddisc}

Here we record the coefficient of $\log^2v$ in the one-loop correlator, as obtained from the squares of
anomalous dimensions in eq.~(\ref{one-loop gamma squared}), and normalized as in eq.~(\ref{dDisc2}):
\be
 z\zb(\zb-z) \GG^{(2)}(u,v)\Big|_{\log^2v} =D(D-2)\bar D (\bar D-2) \GG^{(2)'}(z,\zb) \label{dDisc2appendix}
\ee
where:
\ba
 \GG^{(2)'}(z,\zb) &=& R_0(z,\zb) + R_1(z,\zb)\left( \log z - \log \zb \right)+R_2(z,\zb)\left( \log z + \log \zb \right) \\
 & & + R_3(z,\zb) \left(\Li_2(1-z)-\Li_2(1-\zb) \right)+ R_4(z,\zb) \left(\Li_2(1-\frac{1}{z})-\Li_2(1-\frac{1}{\zb}) \right) \nonumber \\
 & & + \frac{1-\zb}{8u} \Li_2(1-z) - \frac{1-z}{8u} \Li_2(1-\zb) \nonumber
\ea
where
\ba 
R_0(z,\zb) &=& \frac{u v^3 (3 v-7 (u+1))}{16 (z-\zb)^5}+\frac{v^2 (-4 u-3 v+15)}{48 (z-\zb)^3}+\frac{v \left(\frac{7 u}{3}-v-3\right)}{16 u (z-\zb)}-\frac{z-\zb}{16 u}\\
R_1(z,\zb) &=& \frac{v^2 \left(u^2-u+v-1\right)}{8 (z-\zb)^4}+\frac{u v^3 \left(u^2-u v+5 u-v+1\right)}{8 (z-\zb)^6}+\frac{(1-v) v}{8 u (z-\zb)^2}\\
& & -\frac{(1-u)^2+5 v}{96 (z-\zb)^2}+\frac{v (v+1)-2 (1-u)^2}{64 u v}+\frac{13}{192} \\
R_2(z,\zb) &=& \frac{u \left(1-u^2\right) v^2}{8 (z-\zb)^5}+\frac{v}{8 u (z-\zb)}+\frac{v (u (1-u)-6 (-u+v+1))}{96 (z-\zb)^3}\\
& & +\frac{(2 u-v-2) (z-\zb)}{64 u v}+\frac{1-u+v}{96 (z-\zb)}\\
R_3(z,\zb) &=& \frac{u v^2 (u-v-1)}{8 (z-\zb)^6}+\frac{v (u-v-1)}{8 u (z-\zb)^2}+\frac{v^2}{4 (z-\zb)^4}\\
R_4(z,\zb) &=& \frac{u^3 v^2 (u+v-1)}{8 (z-\zb)^6}
\ea

\section{Twist conformal blocks}
\label{apTCB}

In the body of the note we have introduced the following family of functions denoted twist conformal blocks (TCB)\footnote{The conventions used in this paper are slightly different to the ones used in \cite{Alday:2016njk} but better suited for our current purposes.}

\begin{equation}
H_n^{(m)}(z,\bar z)= \sum_\ell \frac{g_{n,\ell}(z,\bar z)}{J^{2m}}\,.
\end{equation}
where the super-conformal blocks have been given in the body of the paper.  We will be interested in the divergent contribution to TCB as $\bar z \to 1$. Given the specific structure of the super-conformal blocks this contribution admits the following factorised form 
\begin{equation}
H_n^{(m)}(z,\bar z)= \frac{1}{z\zb (\zb-z)}  r_{h_n}k_{h_n}(z) \bar H_n^{(m)}(\zb) + \text{regular}
\end{equation}
where recall $h_n=n+3$ and $r_{h_n}$ and $k_{h_n}(z)$ have been defined in the body of the paper. Regular terms behave as a finite number of conformal blocks as $\bar z \to 1$ and will not be important for our discussion. The functions $\bar H_n^{(m)}(\zb) $ satisfy the following recursive relation
\begin{equation}
\bar D \bar H_n^{(m)}(\zb) =\bar H_n^{(m)}(\zb),
\end{equation}
with $\bar D=\bar z^2 \partial_{\bar z} (1-\bar z)\partial_{\bar z}$. From the explicit form of the conformal blocks, or from the correlator at zeroth order, we can compute the divergent behaviour for $\bar H_n^{(0)}(\zb)$. We obtain
\begin{equation}
\bar H_n^{(0)}(\zb)= \frac{1}{2}\frac{1}{1-\bar z} + \text{regular}
\end{equation}
Due to our definition of TCB, which differs from that of \cite{Alday:2016njk}, we see that the divergent part of the TCB is actually independent of $n$, and hence the index will be dropped from now on. In order to obtain the divergence behaviour of the correlator at zeroth order we need to insert $\langle a^{(0)}\rangle_{n,\ell}=2(J^2-(n+2)(n+3))$. This can be obtained by acting with $2(\bar D-(n+2)(n+3))$ on the function above and reinstates the $n$ dependence. From now on we will also drop regular terms. All equalities must be understood up to those. Starting from $\bar H^{(0)}(\zb)$ we can build all functions $\bar H^{(m)}(\zb)$ by using the recursion relations (for instance as an expansion around $\bar z=1$). For $m=1,2,\cdots$ the structure is as follows
 \begin{eqnarray}
\bar H^{(m)}(\zb) =q^{(m)}(\bar z) \log^2(1-\bar z),~~~m=1,2,\cdots
\end{eqnarray}
where
\begin{equation}
\bar D q^{(m+1)}(\bar z)=q^{(m)}(\bar z),~~~~q^{(1)}(\bar z) =\frac{1}{4}
\end{equation}
and $q^{(m)}(\bar z) \sim (1-\bar z)^{m-1}$ as $\bar z \to 1$. It is straightforward to compute the functions $q^{(m)}(\bar z)$ as a series around $\bar z=1$. In the body of the paper we will be interested in reproducing the divergence $(1-\bar z)^{-1}$: 
\begin{equation}
\sum_{m=0} \alpha_m \bar H^{(m)}(\zb) = \frac{1}{1-\bar z}
\end{equation}
From the expressions above it is clear $\alpha_0=2$. Furthermore, the behaviour for $q^{(m)}(\bar z)$ around $\bar z=1$ implies $\alpha_m=0$ for $m>0$.

\subsection{From large spin perturbation theory to an inversion integral}

We are interested in solving the following inversion problem. Given a divergent function, how do we write it in terms of the basis $\bar H^{(m)}(\zb)$. Let us first assume the divergence is proportional to $\log^2 (1-\bar z)$ and the function in front admits a decomposition in terms of $q^{(m)}(\bar z)$.  Later on we will relax this assumption. The first step is to construct the following projectors
\begin{equation}
\oint_{\bar z=1} d\bar z P^{(m)}(\bar z) q^{(m')}(\bar z) = \delta_{m,m'}
\end{equation}
for $m=1,2,\cdots$. The recursion relations for $q^{(m)}(\bar z)$ together with integration by parts lead to the following relation for the projectors
\begin{equation}
P^{(m+1)}(\bar z) = \bar D^\dagger P^{(m)}(\bar z),~~~~P^{(1)}(\bar z)=-\frac{4}{\bar z(1-\bar z)}
\end{equation}
where $\bar D^\dagger = \partial_{\bar z} (1-\bar z)\partial_{\bar z} \bar z^2$. Note a nice point. Now $P^{(m+1)}(\bar z)$ can be easily computed from $P^{(m)}(\bar z)$, by simply acting with a differential operator. Then, given the expansion
\begin{equation}
\sum_{m=1} c_m  q^{(m)}(\bar z) = g(\bar z)
\end{equation}
the coefficients $c_m$ admit the following integral representation
\begin{equation}
c_m = \oint_{\bar z=1} d\bar z P^{(m)}(\bar z)g(\bar z)
\end{equation}
Having found an integral expression for the coefficients $c_m$ we are interested in the following resumed series:
\begin{equation}
f(J)= \sum_{m=1} \frac{c_m}{J^{2m}}
\end{equation}
For this we need to compute the following Kernel
\begin{equation}
K(J,\bar z) = \sum_{m=1} \frac{P^{(m)}(\bar z)}{J^{2m}}
\end{equation}
which plays the role of the generating function for the projectors $P^{(m)}(\bar z)$. Remarkably, one is able to find a closed form expression for this. We find 
\begin{equation}
\label{Kdef}
K(J,\bar z) = -4 \sum_m^\infty \Gamma^2(m+1) \frac{1}{\prod_{k=0}^m(J^2-k(k+1))} \frac{1}{\bar z(1-\bar z)}\left(\frac{\bar z}{1-\bar z} \right)^m
\end{equation}
With this we can give 
\begin{equation}
f(J)= \oint_{\bar z=1} d\bar z K(J,\bar z) g(\bar z)
\end{equation}
In other words, we have found an integral expression for $f(J)$ which is analytic in the spin, and which can be interpolated down to finite spin! very much as in the Froissart-Gribov inversion formula. Let us study this Kernel in greater detail. It admits several representations. One in terms of hypergeometric functions $~_3F_2$. Another in terms of an infinite sum of standard hypergeometric functions $~_2F_1$, by noticing that the Kernel satisfies the differential equation
\begin{equation}
\label{difeq}
\left( \bar D^\dagger  - J^2 \right) K(J,\bar z) = \frac{4}{\bar z(1-\bar z)}
\end{equation}
The most convenient representation for us is as a double integral:
\begin{equation}
K(J,\bar z) = \frac{4}{\bar z(1-\bar z)} \int_0^1 dt_1 \int_0^\infty dt_2 \frac{(1-t_1)^{\hb-1}(1+t_2)^{-\hb}}{t_1 t_2 \zeta-1},~~~~~~\zeta = \frac{\bar z}{1-\bar z}
\end{equation}
where we have defined $J^2=\hb(\hb-1)$. One can explicitly check that the large $J$ expansion coincides with that of (\ref{Kdef}) and furthermore that the differential equation (\ref{difeq}) is satisfied. Plugging this into $f(J)$ and deforming the contour we can integrate around the pole at $\bar z + t_1 t_2 \bar z -1 =0$. We obtain
\begin{equation}
\label{LSPTinversion}
f(J)=4 \int_0^1 dt_1 \int_0^\infty dt_2 (1-t_1)^{\hb-1}(1+t_2)^{-\hb} g\left(\frac{1}{1+t_1 t_2}\right)
\end{equation}
which is a very neat expression, analytic in the spin. So far we have considered the case of a divergence proportional to $\log^2(1-\bar z)$. More generally, we need to consider the double-discontinuity ${\rm dDisc}\,G(\bar z)$. For divergences containing a $\log^2(1-\bar z)$, this reduces to what we did before. Let us now consider a divergence proportional to a non-integer or negative power of $(1-\bar z)$. In this case we obtain
\begin{equation}
4 \int_0^1 dt_1 \int_0^\infty dt_2 (1-t_1)^{\hb-1}(1+t_2)^{-\hb} \frac{1}{4\pi^2} \left. {\rm dDisc}\left(\frac{1-\bar z}{\bar z} \right)^p \right|_{\bar z =\frac{1}{1+t_1 t_2}} =  2 \frac{\Gamma(h-p-1)}{\Gamma^2(-p)\Gamma(h+p+1)}
\end{equation}
One can explicitly check that this gives the precise large $J$ expansion for a divergence corresponding to non-integer or negative $p$. This is relevant to compute the contribution to the CFT data due to an exchanged operator, in the dual channel, of arbitrary twist. This case was analysed in \cite{Alday:2015ewa} for leading twist operators from the point of view of large spin expansions. The formula above fully reproduces those results. 

Before proceeding Let us make an important remark regarding different representations for the Kernel (\ref{Kdef}). Although all representations give the correct large $J$ expansion for integer $p$, or for a divergence proportional to $\log^2(1-\bar z)$, the representation we have used extends properly to non integer $p$. This can be understood intuitively as follows. Other representations involve the integral over $\bar z$ from one to infinity. While this is well defined for integer $p$ it is not for general $p$. The representation we have used is well defined for all $p$ and one can check that indeed gives the correct answer in all examples. 

\section{Results for OPE coefficients}
\label{app:ope}

Here we discuss a simple trick to deal with the $\zb$ integrations of the logarithms and polylogarithms in eq.~(\ref{dDisc2appendix}),
order by order in the $z$ expansion.
The idea is to expand the polynomials multiplying these in a basis of eigenfunctions of the Casimir; this basis is
provided by $k$ functions with negative argument, $K_m(\zb)= \frac{(2m)!}{(m!)^2}k_{-m}(\zb)$.
By integrating-by-parts the relation $(D_\zb-m(m+1))K_m(\zb)=0$, we get an interesting result:
\be
 \int_0^1 \frac{d\zb}{\zb^2} \frac{r_\hb}{\hb-1/2} k_\hb(\zb) K_m(\zb)  = \frac{2}{J^2-m(m+1)}
\ee
where the right-hand-side stems from a boundary term at $\zb=1$.  This formula can be easily checked against the general integral eq.~(\ref{Iint1})
for specific polynomials.
Most importantly, the same method can be applied to polynomials times log, for example:
\be
 \int_0^1 \frac{d\zb}{\zb^2} \frac{r_\hb}{\hb-1/2} k_\hb(\zb) K_m(\zb) \log(\zb) = \int_0^1 \frac{d\zb}{\zb^2} \frac{r_\hb}{\hb-1/2}k_\hb(\zb)
 \frac{\big(D_\zb -m(m+1)\big)(K_m(\zb)\log \zb)}{J^2-m(m+1)}
\ee
where the right-hand-side is just a polynomial in $1/\zb$, which we already know how to deal with.
The same trick, applied repeatedly, takes care of the other transcendental functions which appear in the one-loop
double-discontinuity (\ref{dDisc2appendix}), namely $\Li_2(1-\zb)$ and $\log^2\zb$.
In these cases, however, one finds a new integral: one with a single positive power of $\zb$.
Using the integral representation for the hypergeometric, we find that it integrates to an harmonic sum:
\be
 \int_0^1 \frac{d\zb}{\zb^2} \frac{r_\hb}{\hb-1/2} k_\hb(z) \zb = -4\int_0^1 \frac{dx\,x^{\hb-1}}{1+x}\log x = 4\psi'(\hb)-2\psi'(\frac{1+\hb}{2}) \equiv 4 \bar S_{-2}(\hb-1)
\ee
where $\psi(x)=(\log\Gamma(x))'$ is the polygamma function. This can be regarded as an analytic continuation
of an harmonic sum: $\bar S_{-a}(j) = S_{-a}(j)- S_{-a}(\infty)$ from even $j$, where
\begin{equation}
S_{-a}(j)= \sum_{m=1}^j \frac{(-1)^m}{m^a}\,.
\end{equation} 
It is worth mentioning the large $J$ expansion of $\bar S_{-2}(\hb-1)$ respect reciprocity and contains only powers of $1/J^2$,
where $J^2=\hb(\hb-1)$.  In terms of this special function, our result is
\begin{eqnarray*}
S_{0,\hb}^{(0)}&=& \frac{48}{J^4}+\frac{31672296}{25025 \left(J^2-56\right)}-\frac{169064}{195 \left(J^2-30\right)}-\frac{5490}{J^2-20}-\frac{9720}{\left(J^2-20\right)^2}+\frac{2480}{33 \left(J^2-12\right)} \nonumber \\
& & +\frac{121154}{25 \left(J^2-6\right)}-\frac{624}{\left(J^2-6\right)^2}-\frac{7200}{\left(J^2-6\right)^3}+\frac{306}{J^2-2}+\frac{648}{\left(J^2-2\right)^2}+\frac{3642}{35 J^2} \nonumber \\
& & + \left( \frac{1120}{13 \left(J^2-30\right)}-\frac{80}{11 \left(J^2-12\right)}+\frac{96}{J^2-6}-\frac{18144}{143 \left(J^2-56\right)} \right)\pi^2  \\
 &  & + \left(  \frac{13440}{13 \left(J^2-30\right)}-\frac{960}{11 \left(J^2-12\right)}-\frac{217728}{143 \left(J^2-56\right)}\right) \bar S_{-2}(\hb-1)\,,\\
 S_{1,\hb}^{(0)}&=& -\frac{3000}{J^4}+\frac{41435928}{1001 \left(J^2-72\right)}-\frac{381416}{11 \left(J^2-42\right)}-\frac{211050}{J^2-30}-\frac{693000}{\left(J^2-30\right)^2}+\frac{5175000}{1001 \left(J^2-20\right)}\\
 & & +\frac{432758}{J^2-12}+\frac{240240}{\left(J^2-12\right)^2}-\frac{2469600}{\left(J^2-12\right)^3}-\frac{15150732}{77 \left(J^2-6\right)}+\frac{295200}{\left(J^2-6\right)^2}+\frac{1080000}{\left(J^2-6\right)^3}\\
 & & -\frac{218970}{7 \left(J^2-2\right)}-\frac{51840}{\left(J^2-2\right)^2}-\frac{6850}{J^2} \\
 & & + \left(\frac{39200}{11 \left(J^2-42\right)}-\frac{540000}{1001 \left(J^2-20\right)}+\frac{16800}{J^2-12}-\frac{1108400}{77 \left(J^2-6\right)}-\frac{604800}{143 \left(J^2-72\right)} \right) \pi^2\\
 & & + \left(  -\frac{470400}{11 \left(J^2-42\right)}+\frac{6480000}{1001 \left(J^2-20\right)}-\frac{4800}{77 \left(J^2-6\right)}+\frac{7257600}{143 \left(J^2-72\right)}\right) \bar S_{-2}(\hb-1)\,.
\end{eqnarray*}
Results for higher twists are available upon request to the authors; algorithmically, it is relatively straightforward to obtain similar formulas for the first few hundreds of twists.

\section{Ten-dimensional box integral}
\label{boxes}

Here we record the result for the box integral in ten-dimensions, obtained using e.g. dimensional regularization:
\ba
 I_{box}(s,t) &\equiv& \int \frac{d^{10}p}{i\pi^5} \frac{1}{p^2(p-p_1)^2(p-p_1-p_2)^2(p+p_4)^2}
\nl &=& \frac{1}{120} \Big( \frac{s^2t^2}{u^3} \left(\log^2\frac{-s}{-t}+\pi^2\right)-(s-t)\left(\frac{st}{u^2}+\frac12\right)\log\frac{-s}{-t}
+ u \log \frac{\sqrt{-s}\sqrt{-t}}{\Lambda^2}-\frac{st}{u}\nl
&& \qquad\quad +C_1 \Lambda^2+C_2 u\Big)
\label{Ibox}
\ea
where $\Lambda$ is a ultraviolet cutoff and where we have re-instated the quadratic divergence $C_1$.
$C_2$ is a scheme-dependent constant which is unimportant in this paper (it cancels out when summing the three boxes).
Following standard notation, all logarithms are real in the Euclidean region $s,t<0$ and one adds
a small imaginary part, $-s\mapsto -s-i0$, to select the correct branch when an invariant becomes timelike.
Explicitly, adding up the three boxes and paying due care to phases,
we obtain the function of angles $f_2(x)$ entering eq.~(\ref{A5_sugra}) in the main text,
where $x=-u/s=\frac{1+\cos\theta}{2}$:
\ba \label{f2}
 f_2(x)&=&\frac{1}{3840}\Bigg\{
\Bigg[\frac{x}{x-1}+\frac{x^2}{(x-1)^3}\log x (\log x +2\pi i)
   + x^2\left(\frac{x-3}{(1-x)^2}+3-2x\right)\big(\!\log x+i\pi\big)
\nl&& \hspace{14mm}
   + (x\mapsto1{-}x)\Bigg]
+   x^2(1-x)^2\left(\log \frac{x}{1-x}+\pi^2\right)-x(1-x) + 3C_1\frac{\Lambda^2}{s}\Bigg\}.
\ea

\end{document}